\documentclass[preprint2]{emulateapj}

\input{ulem.sty}
\shorttitle{Spatial Clustering of RASS-AGNs III -- Expanded Sample}
\shortauthors{Krumpe et al.}

\begin{document}

\def\mpch {$h^{-1}$ Mpc} 
\def\kpch {$h^{-1}$ kpc} 
\def\kms {km s$^{-1}$} 
\def\lcdm {$\Lambda$CDM } 
\def\xir {$\xi(r)$}
\def\wprp {$w_p(r_p)$}
\def\xisp {$\xi(r_p,\pi)$}
\def\xis {$\xi(s)$}
\def\rr {$r_0$}
\def\etal {et al.}

\title{The Spatial Clustering of ROSAT All-Sky Survey AGNs\\ III. Expanded Sample and Comparison with Optical AGNs}

\author{Mirko Krumpe\altaffilmark{1,2}, Takamitsu Miyaji\altaffilmark{3,4}, Alison L. Coil\altaffilmark{1,5}, 
  and Hector Aceves\altaffilmark{3}}

\altaffiltext{1}{University of California, San Diego, Center for Astrophysics and
                Space Sciences, 9500 Gilman Drive, La Jolla, CA 92093-0424, USA}
\altaffiltext{2}{European Southern Observatory, ESO Headquarters, 
                 Karl-Schwarzschild-Stra\ss e 2, 85748 Garching bei M\"unchen, Germany}
\altaffiltext{3}{IAUNAM-E (Instituto de Astronom\'ia de la Universidad Nacional
                 Aut\'onoma de M\'exico, Ensenada), PO Box 439027, San Diego, 
                 CA, 92143-9027, USA}
\altaffiltext{4}{Visiting Scholar, University of California, San Diego, 
                 Center for Astrophysics and Space Sciences, 9500 Gilman 
                 Drive, La Jolla, CA 92093-0424, USA}
\altaffiltext{5}{Alfred P. Sloan Fellow}
\email{mkrumpe@ucsd.edu}

\begin{abstract}

This is the third paper in a series that reports on our investigation of the clustering properties 
of AGNs identified in the {\it ROSAT} All-Sky Survey (RASS) and Sloan Digital Sky Survey (SDSS). 
In this paper, we extend the redshift range to $0.07<z<0.50$ and measure 
the clustering amplitudes of both X-ray and optically-selected SDSS broad-line AGNs with 
and without radio detections as well as for X-ray selected narrow-line RASS/SDSS AGNs. 
We measure the clustering amplitude through cross-correlation functions (CCFs) with SDSS galaxies 
and derive the bias by applying a halo occupation distribution (HOD) model directly to the CCFs. 
We find no statistically convincing difference in the clustering of X-ray and 
optically-selected broad-line AGNs, as well as with samples in which radio-detected AGNs are 
excluded.
This is in contrast to low redshift optically-selected narrow-line AGNs, where radio-loud 
AGNs are found in more massive halos than optical AGNs without a radio-detection.
The typical dark matter halo masses of our broad-line AGNs are log 
$(M_{\rm DMH}/[h^{-1}\,M_{\odot}]) \sim 12.4-13.4$, consistent with the halo mass range of 
typical non-AGN galaxies at low redshifts. 
We find no significant difference between the clustering of X-ray selected narrow-line AGNs 
and  broad-line AGNs. We confirm the weak dependence of the clustering strength on AGN X-ray 
luminosity at a $\sim$2$\sigma$ level. Finally, we summarize the current picture of AGN clustering 
to $z\sim 1.5$ based on three dimensional clustering measurements. 

\end{abstract}

\keywords{galaxies: active -- X-rays: galaxies -- 
cosmology: large-scale structure of Universe}


\section{Introduction}
\label{introduction}
 
Galaxy clustering measurements of large area surveys, such as the 2dF Galaxy Redshift Survey (2dFGRS, 
\citealt{colless_dalton_2001}), Sloan Digital Sky Survey (SDSS, \citealt{abazajian_adelman-mccarthy_2009}), 
Deep Extragalactic Evolutionary Probe 2 Redshift Survey (DEEP2, \citealt{davis_faber_2003}), 
AGN and Galaxy Evolution Survey (AGES, \citealt{kochanek_eisenstein_2004}), and
VIMOS-VLT Deep Survey (VVDS, \citealt{lefevre_vettolani_2005})
with many thousands 
of galaxies have quantified the clustering dependence on galaxy properties such as 
morphological type, luminosity, spectral 
type, and redshift to $z\sim1$ (e.g., \citealt{norberg_baugh_2002}; \citealt{madgwick_hawkins_2003}; \citealt{zehavi_zheng_2005}; 
\citealt{meneux_fevre_2006}; \citealt{coil_newman_2008}; \citealt{meneux_guzzo_2009}).
The general trend is that red, more massive, brighter, and/or elliptical galaxies cluster more strongly 
than blue, less massive, fainter, and/or spiral galaxies.
Halo occupation distribution (HOD) modelling of these results has shown 
that there is an almost linear increase in the mean number of satellite 
(non-central) galaxies as a function of increasing 
dark matter halo (DMH) mass (e.g., \citealt{zheng_coil_2007}; \citealt{zehavi_zheng_2011}).

Compared to projected angular clustering measurements, three dimensional (3D) clustering measurements 
have greater statistical accuracy and minimize systematic errors associated with model assumptions 
in using Limber's de-projection (\citealt{limber_1954}).
However, 3D correlation functions require extremely large numbers of objects with secure redshift 
information.

Clustering measurements of the auto-correlation function (ACF) of active galactic nuclei (AGNs) reveal 
how AGNs are spatially distributed in the Universe.
These measurements can constrain the AGN host galaxies, determine the 
typical DMH mass in which AGNs reside and the distribution of AGNs with DHM mass, test 
theoretical model predictions, and address which physical processes are triggering AGN activity
(e.g., \citealt{porciani_magliocchetti_2004}; \citealt{gilli_daddi_2005,gilli_zamorani_2009}; 
\citealt{yang_mushotzky_2006}; \citealt{coil_hennawi_2007}; \citealt{coil_georgakakis_2009}; 
\citealt{ross_shen_2009}; \citealt{krumpe_miyaji_2010}; \citealt{cappelluti_ajello_2010}; 
\citealt{miyaji_krumpe_2011}; \citealt{allevato_2011}). However, a major challenge in AGN 
clustering measurements is to 
overcome the limitation caused by the low space density of AGNs. 

Narrow emission line diagnostics (e.g., \citealt{veilleux_osterbrock_1987}; 
\citealt{kauffmann_heckman_2003}) have been used to identify narrow-line AGNs at low redshifts.
Large sky area surveys with extensive spectroscopic follow-up programs have 
recently allowed narrow-line AGN clustering measurements for samples of up 
to 90,000 AGNs 
(\citealt{li_kauffmann_2006}). Studies such as \cite{wake_miller_2004}, \cite{li_kauffmann_2006}, 
\cite{magliocchetti_maddox_2004}, and \cite{mandelbaum_li_2009} explore the 
clustering strength of low redshift ($z<0.3$) narrow-line AGNs samples with 
respect to various AGN properties (e.g., black hole mass, radio emission).
Optically-selected narrow-line AGNs have a clustering strength similar to 
all galaxies (both blue and red). Narrow-line radio-loud AGNs 
cluster more strongly than AGN without radio-emission and identical samples of 
quiescent galaxies. Consequently, denser environment of galaxies increase the probability 
of hosting a narrow-line radio-loud AGN.

Broad-line AGNs are, in general, more luminous than narrow-line AGNs. Moreover, 
their energy production dominates over the host galaxy star light. Large area 
optical surveys have spectroscopically identified tens of thousands of 
broad-line AGNs (e.g., \citealt{schneider_richards_2010}; \citealt{croom_smith_2001}). 
The co-moving number density of broad-line AGNs is very low in the low redshift universe ($z<0.5$). 
The situation improves at higher redshifts as the co-moving AGN number distribution 
peaks at $z\sim 2$. Furthermore, a given observed area corresponds at 
higher redshift to a larger observed 
co-moving volume. These two facts favor broad-line AGN clustering measurements at 
$z \gtrsim 0.5$ (e.g., \citealt{porciani_magliocchetti_2004}; \citealt{croom_boyle_2005}; 
\citealt{ross_shen_2009}).

X-ray surveys have also been used to select AGNs. These 
surveys cover much less sky area, e.g., $\sim$0.1-10 deg$^2$, than 
optical surveys.  Further, they 
sample lower AGN luminosities than optical surveys and contain 
a significant fraction of obscured AGNs. In order to probe 
the required large observed 
co-moving volume needed for clustering measurements using these 
relatively small sky areas, 3D clustering measurements of 
X-ray selected AGNs have mainly been conducted at $z\gtrsim 0.8$ 
(e.g., \citealt{gilli_daddi_2005,gilli_zamorani_2009}; \citealt{yang_mushotzky_2006}; 
\citealt{coil_hennawi_2007}). 

Optically-selected AGNs at redshifts of $z\sim1.0-1.5$ 
appear to reside in somewhat lower host DMH masses 
($M_{\rm DMH} \sim 10^{12-13}\,h^{-1}$ M$_{\odot}$, e.g., \citealt{porciani_magliocchetti_2004}; 
\citealt{coil_hennawi_2007}; \citealt{ross_shen_2009}) than {\it XMM-Newton} and {\it Chandra}
selected AGN samples ($M_{\rm DMH} \sim 10^{13-13.5}\,h^{-1}$ M$_{\odot}$, 
e.g., \citealt{yang_mushotzky_2006}; \citealt{gilli_daddi_2005,gilli_zamorani_2009}; 
\citealt{coil_georgakakis_2009}; \citealt{allevato_2011}). Possible 
explanations of the differences in the clustering signals include either the presence of a large 
fraction of X-ray absorbed narrow-line type II AGNs in the {\it XMM-Newton} and {\it Chandra} 
samples or the different luminosities of the AGN samples.
At lower redshifts 3D broad-line X-ray 
AGN clustering measurements have been associated 
with large uncertainties, due to small sample sizes 
(e.g., \citealt{mullis_henry_2004}).

To achieve smaller uncertainties in AGN clustering measurements, 
\cite{coil_georgakakis_2009} use the cross-correlation function (CCF) of AGNs with a 
tracer set of galaxies that contains a large number of objects. The AGN ACF is then inferred from the CCF,  
which has many more pairs at a given separation and hence significantly reduces the 
uncertainties in the spatial correlation function compared with the direct measurements of the AGN ACF. 

In \cite{krumpe_miyaji_2010} (hereafter paper I) we use the cross-correlation 
technique to calculate the CCF between broad-line  X-ray selected AGNs from the {\em ROSAT} All 
Sky Survey (RASS) and SDSS Luminous Red Galaxies (LRGs) in the redshift range $0.16<z<0.36$. 
The potential of RASS, which is currently the most sensitive X-ray all sky survey, 
can only be maximally exploited when it is combined with other large-area surveys such as SDSS. 
The unprecedented low uncertainties of the inferred broad-line AGN ACF from the RASS/SDSS combination 
allows us to split our sample into subsamples according to their X-ray luminosities. From this work,
for the first time, we report an X-ray luminosity dependence of broad-line AGN clustering in that 
higher luminosity AGNs cluster more strongly than their lower luminosity counterparts. We conclude that 
low luminosity broad-line RASS/SDSS AGNs cluster 
similarly to blue galaxies at the same redshift, while high luminosity RASS/SDSS AGNs cluster similarly 
to red galaxies.  

In our second paper (\citealt{miyaji_krumpe_2011}; hereafter paper II), we apply a halo occupation 
distribution (HOD) modeling technique to the AGN-LRG CCF in order to move beyond determining the 
typical DMH mass based on the clustering signal strength and instead constrain the full 
distribution of AGNs as a function of DMH mass. To do this, we develop a novel method of applying the
HOD model directly to the CCF. The HOD modeling significantly improves the analysis of the CCF because 
it properly uses the Fourier transformed linear power spectrum in the  
``two halo term'' as well as the non-linear growth of matter in the ``one halo term''
through the formation and growth of DMHs. This results in significant improvements
over the standard method, which is used in 
paper I, of fitting both regimes with a phenomenological power law. One of the important results of 
this analysis is that at $0.16<z<0.36$ the mean number of satellite AGNs in a DMH does not proportionally 
increase with halo mass, as is found for satellite galaxies. The AGN fraction 
among satellite galaxies actually {\it decreases} with increasing DMH mass beyond 
$M_{\rm DMH} \sim 10^{12}$ $h^{-1}$ M$_{\odot}$.

In this paper we extend the scope of our previous papers to both somewhat lower and higher 
redshifts to obtain broad-line AGN clustering results at $z<0.5$, where 
very precise narrow-line AGN clustering measurements exist but broad-line AGN clustering is poorly 
constrained. This is crucial for studying the possible evolution and luminosity dependence of 
broad-line AGN clustering from low to high redshifts. The dominant process that triggers
AGN activity could be a function of redshift and/or halo mass, which may be reflected
in the clustering properties. We also study the clustering signal of both X-ray and optically-selected 
broad-line AGN samples, and test whether the exclusion of radio-detected broad-line AGNs changes our results.  
As the same statistical method and galaxy tracer sets are used to infer the 
clustering signal for X-ray and optically-selected broad-line AGNs, we can explore differences 
in the clustering among the different selection techniques with low systematic uncertainties.
Furthermore, we derive bias parameters by applying the HOD modeling directly to all CCFs.
In Paper II we show that using power law fit results even in the non-linear regime, as is 
commonly done in the literature, is appropriate to detect difference in the clustering properties
between different samples. However, the derived bias parameters and DMH masses based 
on these fits should be interpreted with caution as the fit does not only consider the linear 
regime (two halo term), but also the non-linear regime (one halo term).
Consequently, here we derive the bias for each AGN sample using HOD modeling of the CCF.  
Full detailed results of the HOD modeling of the CCFs presented in this paper will be given in 
a future paper (Miyaji et al. in preparation).

This paper is organized as follows. In Section 2 we describe the properties of the 
different galaxy tracers sets used at different redshifts, while Section 3 gives the details of 
our different AGN samples. In Section 4 we briefly summarize the cross-correlation technique, 
how the AGN ACF is inferred from this, and present our results. We 
apply the HOD modeling in Section 5. Our results are discussed in Section 6 and we conclude 
in Section 7. Throughout the paper, all distances are measured in co-moving 
coordinates and given in units of $h^{-1}$\,Mpc, where $h= H_{\rm 0}/100$\,km\,s$^{-1}$\,Mpc$^{-1}$, unless 
otherwise stated. We use a cosmology of $\Omega_{\rm m} = 0.3$, $\Omega_{\rm \Lambda} = 0.7$, and 
$\sigma_8(z=0)=0.8$, which is consistent with the $WMAP$ data release 7 
(\citealt{larson_dunkley_2011}; Table~3). The same cosmology is used in papers I \& II.
Luminosities and absolute magnitudes are calculated for $h=0.7$.
We use AB magnitudes throughout the paper. All uncertainties represent to a 1$\sigma$ 
(68.3\%) confidence interval unless otherwise stated.


\section{Galaxy Tracer Sets}

The crucial ingredient in the cross-correlation method is the tracer set, a sample with a large number 
of objects that traces the underlying dark matter density distribution. The properties of the tracer
set determines the redshift range over which the method can be applied. The AGN samples of interest 
are necessarily limited to the same redshift range and geometry as the corresponding tracer set. 
As the RASS/SDSS selected AGNs (\citealt{anderson_margon_2007}) are based on the SDSS data release 5 (DR5), 
we consequently limit the tracer sets to the same survey geometry when we compute the 
clustering measurements of the X-ray selected AGN. The SDSS geometry and completeness are 
expressed in terms of spherical polygons (\citealt{hamilton_tegmark_2004}). This file is 
not publicly-available for DR5, therefore, we use the latest version available prior to DR5: 
the DR4+ geometry file\footnote{\tt http://sdss.physics.nyu.edu/lss/dr4plus}. However, optically-selected 
SDSS AGNs (\citealt{schneider_richards_2010}) make use of the full SDSS survey (DR7). 
Consequently, we consider tracer sets from the DR7 
geometry\footnote{\tt http://sdss.physics.nyu.edu/lss/dr72} whenever we compute CCFs of 
optically-selected SDSS AGNs.  

In the redshift range of $0.07<z<0.16$ we use SDSS main galaxies for the tracer set, while SDSS 
luminous red galaxies (LRGs, \citealt{eisenstein_annis_2001}) are used for the corresponding 
cross-correlation sample at $0.16<z<0.36$ (same as for paper I). Very luminous red galaxies are 
used as a tracer set at $0.36<z<0.50$. We will refer to the latter sample as the 'extended LRG 
sample'. Above 
$z \sim 0.5$ the number of galaxies with spectroscopic redshifts in SDSS decreases 
dramatically and does not allow the selection of further tracer sets with a sufficient density  
of objects. In the following subsections we describe in detail the extraction of the various tracer 
sets and how we account for SDSS fiber collisions.


\subsection{SDSS Main Galaxy Sample\label{MainGal}}

The SDSS Main Galaxy sample (\citealt{strauss_weinberg_2002}) is drawn from the NYU Value-Added Galaxy 
catalog\footnote{{\tt http://sdss.physics.nyu.edu/vagc/}} (NYU VAGC, \citealt{blanton_schlegel_2005}; 
\citealt{padmanabhan_schlegel_2008}), based on the SDSS DR7 (\citealt{abazajian_adelman-mccarthy_2009}).

The photometric data  
covers an area of 10417 deg$^2$, while the spectroscopic data covers 7966 deg$^2$.
Absolute magnitudes, based on Petrosian fluxes, are $K$-corrected to $z=0.1$ (\citealt{blanton_brinkmann_2003}; \citealt{blanton_roweis_2007}), which is close to the median redshift of our sample.
We follow a scheme similar to \cite{zehavi_zheng_2005}, who use 
the NYU VAGC to measure the clustering of various luminosity and color-selected galaxy subsamples. 
Similarly, we limit our sample to $14.5 < r < 17.5$. The bright limit avoids incompleteness due to galaxy 
deblending, and the faint limit accounts for the slightly-varying $r$-band magnitude limit over the 
SDSS area (nominal value $r\sim17.7$). The restriction of $r < 17.5$ ensures a uniform 
flux limit throughout the whole SDSS survey. In addition, we create a volume-limited galaxy 
sample by selecting objects with an absolute magnitude of $-22.1 < M^{0.1}_r < -21.1$. Finally
we limit the redshift range to $0.07 < z <0.16$. 

Applying the above-mentioned selection criteria and considering only SDSS DR7 areas with a 
spectroscopic completeness ratio of $>$0.8, we select 68273 galaxies with spectroscopic redshifts 
from the corresponding NYU VAGC.
The DR7 area covered by restricting the spectroscopic completeness ratio to $>$0.8 is 
7670 deg$^2$. The properties of this DR7 SDSS main galaxy sample are summarized in Table~\ref{samples}.

As described above, the X-ray selected RASS/SDSS AGN samples are based on DR5. Therefore, we further 
reconfigure the DR7 tracer sets (in this case the SDSS main galaxy sample) to the DR4+ geometry 
to define a common survey geometry to use when measuring the clustering of X-ray selected AGNs. 
The restriction to the DR4+ survey area with a DR7 spectroscopic 
completeness ratio of $>$0.8 corresponds to an area of 5468 deg$^2$. Table~\ref{samples} lists the 
properties of the DR4+ SDSS Main Galaxy sample.

\subsubsection{Accounting for the SDSS Fiber Collision}
An operational constraint of the SDSS spectroscopic program is that two fibers cannot 
be placed closer than 55 arcsec on a single plate. Overlapping spectroscopic plates compensate partially 
for the effect. However, $\sim$7\% of the target galaxies cannot be spectroscopically 
observed because of fiber collisions. This
observational bias is corrected by assigning to each galaxy that has not been observed
the redshift of their nearest neighbor with a spectroscopic SDSS 
redshift (\citealt{blanton_schlegel_2005}).

Although one might be concerned that this simple method could overcorrect and result in too many close galaxy 
pairs at the same redshift, which would then distort clustering 
measurements, \cite{zehavi_zheng_2005} demonstrate that this correction procedure works 
very well. They use $\Lambda$CDM $N$-body simulations and design three galaxy samples and measure 
the correlation function for three samples: i) from 
the full simulated galaxy distribution, ii) from simulated SDSS data including fiber collision losses 
and not correcting for it, and iii) from simulated SDSS data that corrects for the fiber collision by assigning the 
redshift of their nearest spectroscopic neighbor.
They verify that the differences between i) and iii) are much smaller than the statistical uncertainties 
down to scales of $r_{\rm p}\sim 0.1$ $h^{-1}$ Mpc, while ii) underestimates the correlation function at scales 
$r_{\rm p} < 1$ $h^{-1}$ Mpc.
Therefore, we use the same fiber correction procedure for our main galaxy sample. 
The NYU VAGC\footnote{{\tt http://sdss.physics.nyu.edu/vagc-dr7/vagc2/kcorrect/\\kcorrect.nearest.petro.z0.10.fits}}
provides this information through an SDSS fiber collision corrected galaxy sample.

\begin{deluxetable*}{cccccccc}
\tabletypesize{\normalsize}
\tablecaption{Properties of the SDSS Galaxy Tracer Sets and the AGN Samples.\label{samples}}
\tablewidth{0pt}
\tablehead{
\colhead{Sample} &\colhead{SDSS} & \colhead{}  & \colhead{$M,\,$log$L_{\rm X}$ range} & \colhead{Sample} & \colhead{$\langle$$n$$\rangle$} & \colhead{} & \colhead{$\langle$$M,\,$log$L_{\rm X}$$\rangle$}\\
\colhead{Name} &\colhead{Geometry}& \colhead{$z$-range}  & \colhead{(mag,\,erg\,s$^{-1}$)} & \colhead{Size} & \colhead{($h^{3}$ Mpc$^{-3}$)} & \colhead{$\langle$$z$$\rangle$} & \colhead{(mag,\,erg\,s$^{-1}$)}}
\startdata
\multicolumn{8}{c}{SDSS Tracer Sets}\\
Main galaxy & DR7 & $0.07 <z<0.16$ & $-22.1 < M_{r}^{0.1} < -21.1$  & 68273 & $9.8 \times 10^{-4}$ & 0.13& -21.41\\
  sample    & DR4+& $0.07 <z<0.16$ & $-22.1 < M_{r}^{0.1} < -21.1$  & 48994 & $9.8 \times 10^{-4}$ & 0.13& -21.41\\
            &          &                &                                &       &                      &     & \\ 
LRG sample  & DR7 & $0.16 <z<0.36$ & $-23.2 < M_{g}^{0.3} < -21.2$  & 65802 & $9.8 \times 10^{-5}$ & 0.28& -21.71 \\
            & DR4+& $0.16 <z<0.36$ & $-23.2 < M_{g}^{0.3} < -21.2$  & 45899 & $9.6 \times 10^{-5}$ & 0.28& -21.71 \\
            &          &                &                                &       &                      &     & \\ 
Extended LRG& DR7 & $0.36 <z<0.50$ & $-23.2 < M_{g}^{0.3} < -21.7$  & 28781 & $3.9 \times 10^{-5}$ & 0.42& -22.04 \\
    sample  & DR4+& $0.36 <z<0.50$ & $-23.2 < M_{g}^{0.3} < -21.7$  & 19831 & $3.8 \times 10^{-5}$ & 0.42& -22.04 \\\hline
\multicolumn{8}{c}{X-ray Selected AGNs -- RASS/SDSS AGNs}\\
total RASS-AGN          & DR4+& $0.07 <z<0.16$ & $43.05 \lesssim $\ log$L_{\rm X} \lesssim 44.12$ &  629 & $5.2 \times 10^{-5}$ & 0.12 & 43.59 \\
total RASS-AGN (rq)     & DR4+& $0.07 <z<0.16$ & $43.04 \lesssim $\ log$L_{\rm X} \lesssim 44.04$ &  504 & $4.5 \times 10^{-5}$ & 0.12 & 43.55 \\
low $L_{\rm X}$ RASS-AGN & DR4+& $0.07 <z<0.16$ & $42.92 \lesssim $\ log$L_{\rm X}\le       43.54$ &  293 & $4.5 \times 10^{-5}$ & 0.11 & 43.25 \\
high $L_{\rm X}$ RASS-AGN& DR4+& $0.07 <z<0.16$ & $43.54 <        $\ log$L_{\rm X} \lesssim 44.27$ &  336 & $7.0 \times 10^{-6}$ & 0.13 & 43.89 \\
low $L_{\rm X}$ RASS-AGN (rq)& DR4+& $0.07 <z<0.16$&$42.93 \lesssim $\ log$L_{\rm X}\le     43.54$ &  253 & $4.0 \times 10^{-5}$ & 0.11 & 43.26 \\
high $L_{\rm X}$ RASS-AGN (rq)& DR4+&$0.07 <z<0.16$&$43.54 <      $\ log$L_{\rm X} \lesssim 44.16$ &  251 & $5.2 \times 10^{-6}$ & 0.13 & 43.85 \\
narrow line RASS-AGN    & DR4+& $0.07 <z<0.16$ & $42.81 \lesssim $\ log$L_{\rm X} \lesssim 43.81$ &  194 & $6.5 \times 10^{-4}$ & 0.11 & 43.32 \\
              &         &                &                                &       &                      &     & \\    
total RASS-AGN          & DR4+& $0.16 <z<0.36$ & $43.69 \lesssim $\ log$L_{\rm X} \lesssim 44.68$ & 1552 & $6.0 \times 10^{-5}$ & 0.25 & 44.17 \\
total RASS-AGN (rq)     & DR4+& $0.16 <z<0.36$ & $43.69 \lesssim $\ log$L_{\rm X} \lesssim 44.63$ & 1337 & $4.9 \times 10^{-5}$ & 0.25 & 44.15 \\
low $L_{\rm X}$ RASS-AGN & DR4+& $0.16 <z<0.36$ & $43.62 \lesssim $\ log$L_{\rm X} \le      44.29$ &  990 & $5.8 \times 10^{-5}$ & 0.24 & 43.94 \\
high $L_{\rm X}$ RASS-AGN& DR4+& $0.16 <z<0.36$ & $44.29 <        $\ log$L_{\rm X} \lesssim 44.87$ &  562 & $1.2 \times 10^{-6}$ & 0.28 & 44.58 \\
low $L_{\rm X}$ RASS-AGN (rq) & DR4+& $0.16 <z<0.36$ & $43.64 \lesssim $\ log$L_{\rm X} \le  44.29$&  883 & $4.8 \times 10^{-5}$ & 0.24 & 43.95 \\
high $L_{\rm X}$ RASS-AGN (rq)& DR4+& $0.16 <z<0.36$ & $44.29 <    $\ log$L_{\rm X} \lesssim 44.82$&  454 & $1.0 \times 10^{-6}$ & 0.28 & 44.55 \\
narrow line RASS-AGN    & DR4+& $0.16 <z<0.36$ & $43.50 \lesssim $\ log$L_{\rm X} \lesssim 44.40$ &  187 & $7.1 \times 10^{-6}$ & 0.24 & 43.92 \\
             &          &                &                                &       &                      &     & \\   
total RASS-AGN          & DR4+& $0.36 <z<0.50$ & $44.25 \lesssim $\ log$L_{\rm X} \lesssim 45.04$ & 876 & $8.5 \times 10^{-5}$ & 0.43 & 44.64 \\
total RASS-AGN (rq)     & DR4+& $0.36 <z<0.50$ & $44.24 \lesssim $\ log$L_{\rm X} \lesssim 44.99$ & 722 & $8.3 \times 10^{-5}$ & 0.43 & 44.61 \\\hline
\multicolumn{8}{c}{Optically-Selected AGNs -- SDSS AGNs}\\
total SDSS-AGN          & DR7 & $0.07 <z<0.16$ & $-23.26 \lesssim M_{i} \lesssim -22.06$          & 177 & -- & 0.13 & -22.52\\
total SDSS-AGN (rq)     & DR7 & $0.07 <z<0.16$ & $-23.01 \lesssim M_{i} \lesssim -22.06$          & 96  & --  & 0.13 & -22.45\\
             &          &                &                                &       &                      &     & \\    
total SDSS-AGN          & DR7 & $0.16 <z<0.36$ & $-23.27 \lesssim M_{i} \lesssim -22.07$          &3500 &--   & 0.28 & -22.55\\
total SDSS-AGN (rq)     & DR7 & $0.16 <z<0.36$ & $-23.17 \lesssim M_{i} \lesssim -22.06$          &2879 & --  & 0.29 & -22.51\\
total SDSS-AGN (noX)    & DR7 & $0.16 <z<0.36$ & $-23.09 \lesssim M_{i} \lesssim -22.06$          &2367 & --  & 0.29 & -22.47\\
total SDSS-AGN (rq+noX) & DR7 & $0.16 <z<0.36$ & $-22.98 \lesssim M_{i} \lesssim -22.05$          &1958 & --  & 0.29 & -22.44\\
total SDSS-AGN (onlyX)  & DR7 & $0.16 <z<0.36$ & $-23.52 \lesssim M_{i} \lesssim -22.11$          &1133 & --  & 0.27 & -22.72\\
low $M_i$ SDSS-AGN      & DR7 & $0.16 <z<0.36$ & $-22.4       <   M_{i} \lesssim -22.03$          &1757 & --  & 0.28 & -22.18\\
high $M_i$ SDSS-AGN    & DR7 & $0.16 <z<0.36$ & $-23.60 \lesssim M_{i}    \le   -22.4 $          &1743 & --  & 0.29 & -22.93\\
low $M_i$ SDSS-AGN (rq)  & DR7 & $0.16 <z<0.36$ & $-22.4       <   M_{i} \lesssim -22.03$          &1520 &--   & 0.28 & -22.18\\
high $M_i$ SDSS-AGN (rq) & DR7 & $0.16 <z<0.36$ & $-23.52 \lesssim M_{i}    \le   -22.4 $          &1359 & --  & 0.29 & -22.88\\
             &          &                &                                &       &                      &     & \\    
total SDSS-AGN          & DR7 & $0.36 <z<0.50$ & $-23.89 \lesssim M_{i} \lesssim -22.36$          &4404 &--   & 0.43 & -23.04\\
total SDSS-AGN (rq)     & DR7 & $0.36 <z<0.50$ & $-23.79 \lesssim M_{i} \lesssim -22.36$          &3773 &--   & 0.43 & -23.01\\
total SDSS-AGN (noX)    & DR7 & $0.36 <z<0.50$ & $-23.72 \lesssim M_{i} \lesssim -22.35$          &3421 & --  & 0.43 & -22.98\\
total SDSS-AGN (rq+noX) & DR7 & $0.36 <z<0.50$ & $-23.65 \lesssim M_{i} \lesssim -22.35$          &2960 & --  & 0.43 & -22.96\\
low $M_i$ SDSS-AGN      & DR7 & $0.36 <z<0.50$ & $-22.9       <   M_{i} \lesssim -22.22$          &2059 & --  & 0.42 & -22.55\\
high $M_i$ SDSS-AGN     & DR7 & $0.36 <z<0.50$ & $-24.24 \lesssim M_{i}    \le   -22.9 $          &2345 & --  & 0.44 & -23.48\\
low $M_i$ SDSS-AGN (rq) & DR7 & $0.36 <z<0.50$ & $-22.9       <   M_{i} \lesssim -22.24$          &1804 &--   & 0.42 & -22.56\\
high $M_i$ SDSS-AGN (rq)& DR7 & $0.36 <z<0.50$ & $-24.13 \lesssim M_{i}    \le   -22.9 $          &1969 & --  & 0.44 & -23.43

\enddata
\tablecomments{The notation $\langle$$\rangle$ characterizes the average (mean) value of the given quantity. 
               The columns `$M,\,L_{\rm X}$ range' and `$\langle$$M,\,L_{\rm X}$$\rangle$' specify absolute optical magnitudes for 
               optical samples and galactic-absorption corrected 0.1-2.4 keV luminosities for X-ray selected samples. 
               The absolute magnitudes are given for the SDSS filter band stated in the lower index and $K$-corrected to the 
               redshift given in the upper index. The symbol ``$\lesssim$'' is used to characterize the ``soft'' luminosity boundary of the 
               samples; it indicates the 10th (for the lower bound) or the 90th (for the upper bound) percentile. 
               Note that unlike the main galaxy sample and the LRG sample, the 
               extended LRG sample is not volume-limited. The listed co-moving number densities for the extended LRG samples are 
                the number densities for the redshift range $0.36<z<0.42$ and $-23.2<M_g^{0.3}<-21.7$, where the 
                sample is volume-limited. The number densities of the X-ray selected AGN samples are calculated by computing the 
                co-moving volume available to each object.  For optically-selected AGNs, the selection function is 
                complex enough that we do not derive number densities. Abbreviations: rq -- radio-quiet AGNs only (see our 
               definition of radio-quiet in Sections \ref{radio_RASS} and \ref{radio_O_SDSS}); 
                noX -- optically-selected SDSS AGNs that are not detected by RASS; onlyX -- only optically-selected 
                SDSS-AGNs that are also RASS detections.}
\end{deluxetable*}

\subsubsection{Construction of the Random Main Galaxy Sample}
\label{randomcat}
The random sample is another crucial ingredient required for measuring the correlation function. 
The purpose is to create a randomly-distributed sample of objects that 
exactly matches all observational biases (window function, redshift distribution, etc.)
of the observed sample. We follow the procedure of paper I 
(for details see paper I, Section~3.1) and generate a set of random $RA$ and $Dec$ values 
within DR7 areas with a spectroscopic completeness ratio 
of $>$0.8, populate areas with higher completeness ratios more than areas with lower 
completeness ratios, smooth the observed redshift distribution by applying a least-square 
(\citealt{savitzky_golay_1964}) low-pass filter, and use this smoothed redshift 
profile to randomly assign redshifts to the objects in the sample. 

The number of objects in a random catalog is chosen to have an adequate number of 
pairs in the CCF at the smallest scales measured here. For clustering measurements with 
the main galaxy sample the random catalog contains 100 times as many objects 
as the observed sample. 
The random catalog of the DR4+ SDSS main galaxy sample also contains 100 times 
more objects than the data and follows the same exact procedure described 
for the DR7 geometry, except that we restrict that area to the DR4+ geometry.

\subsection{SDSS Luminous Red Galaxy Sample} 
\label{desc_LRG}
The selection of the SDSS LRG sample is described in detail in 
Section~2.1 of paper I. Here we briefly summarize the sample selection. 
We extract LRGs from the web-based SDSS Catalog Archive Server 
Jobs System\footnote{\tt http://casjobs.sdss.org/CasJobs/} using the flag 
'galaxy\_red', which is based on the selection criteria defined in 
\cite{eisenstein_annis_2001}. We verify that the extracted objects meet all LRG 
selection criteria and create a volume-limited sample with $0.16<z<0.36$ and 
$-23.2<M_g^{0.3}<-21.2$, where $M_g^{0.3}$ is based on the extinction-corrected 
$r^{*}_{\rm petro}$ magnitude to construct the $k$-corrected and passively 
evolved rest-frame $g^{*}_{\rm petro}$ magnitudes at $z=0.3$. We consider only  
LRGs that fall into the SDSS area with a DR7 spectroscopic completeness ratio of 
$>$0.8 and have a redshift confidence level of $>$0.95. 

The correction for the SDSS fiber collision in the SDSS LRG sample is slightly 
different from that for the SDSS main galaxy sample. LRGs exhibit very 
well-defined spectra dominated by an old stellar population that evolves very slowly. 
The reduced scatter in the spectral energy 
distribution (SED) of LRGs results in much lower photometric redshift uncertainties than 
the estimates for main galaxies, which can have ongoing star formation and 
therefore have a wider distribution of SEDs. We make use of the precise LRG photometric
redshifts to correct for the SDSS fiber collision in the following manner. 
We select from the SDSS archive all LRGs that pass the pure photometric-based LRG 
selection criteria. We identify photometric LRGs that are closer than 55 arcsec to a 
spectroscopic observed LRG in our redshift and absolute magnitude range. 
We then assign a redshift using the following steps: we accept the 
spectroscopic redshift of the LRG even if its redshift confidence level is 
$\le$0.95. If there is no spectroscopic redshift available for the object, 
we give the photometric LRG the same redshift as the 
spectroscopic neighbor LRG (within a 55 arcsec radius) if  
\begin{equation}\label{eq:redshift}
\mid z_{\rm spec,j} - z_{\rm photo,i} \mid \le \delta z_{\rm photo,i,1\sigma}.  
\end{equation}
If Equation~\ref{eq:redshift} is not fulfilled, we assume that the photometric 
redshift of the LRG is the correct redshift. A redshift is 
assigned only if the object meets the 
selection criteria to construct a volume-limited sample: $0.16<z<0.36$ and 
$-23.2<M_g^{0.3}<-21.2$. Approximately 2\% of the all LRGs in our sample are assigned 
redshifts. The properties of the sample are shown in Table~\ref{samples}. 

The construction of the random catalogs is identical to the procedure described 
in Section~\ref{randomcat} (for details see paper I, Section~3.1), except that 
the LRG random catalogs contain 200 times as many objects as the real 
DR7 and DR4+ LRG samples. This is a compromise between the required 
computation 
time to calculate the correlation functions and having sufficient counts on the 
smallest scales to avoid introducing noise. More objects 
are required at higher redshift 
for a given sky area 
to account for the fact that with increasing redshift the same angular distance 
on the sky corresponds to larger physical co-moving separations.

\begin{figure}
  \centering
 \resizebox{\hsize}{!}{ 
  \includegraphics[bbllx=63,bblly=369,bburx=551,bbury=700]{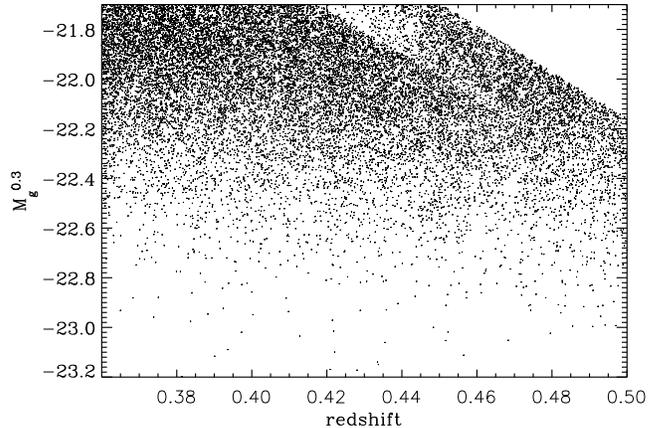}} 
      \caption{Absolute magnitude versus redshift for the extended SDSS 
               luminous red galaxy sample ($0.36 < z< 0.50$). The absolute 
               magnitude is based on the extinction-corrected $r^{*}_{\rm petro}$
               magnitude, passively evolved to $z=0.3$. The different 
               selection criteria for low and high redshift LRGs are visible 
               in the upper right corner (cut I and cut II, see \citealt{eisenstein_annis_2001}).}
         \label{extendedLRG}
\end{figure}

\subsection{Extended SDSS Luminous Red Galaxy Sample}

In order to extend our clustering measurements to higher redshifts, we 
create an ``extended SDSS LRG sample'' over the redshift range of $0.36<z<0.50$.
The extraction of the sample and the correction for the SDSS fiber collision 
problem is identical to the SDSS LRG sample (see Section~\ref{desc_LRG}).

Ideally this sample would also be volume-limited. However, that would require 
an absolute magnitude cut of $M_g^{0.3} = -22.2$ mag, which results in only 
6292 objects. As such a relatively low number of objects would yield 
a measured ACF with very large uncertainties, we use a non-volume-limited
sample with $-23.2<M_g^{0.3}<-21.7$. This is a compromise between making the 
sample volume-limited and retaining accuracy when computing 
the ACF and CCFs (see Table~\ref{samples}). We plot the absolute magnitude versus 
redshift for the 28781 objects (DR7) in the extended LRG sample 
in Fig.~\ref{extendedLRG}.

The co-moving number density for this sample given in Table~\ref{samples} is 
computed over the redshift range $0.36<z<0.42$ and  magnitude range
 $-23.2<M_g^{0.3}<-21.7$, where the sample is 
volume-limited. Furthermore, it assumes that there is no number density 
evolution at higher redshifts 
($0.36<z<0.42$) and in the range  $-23.2<M_g^{0.3}<-21.7$. In principle, we can derive 
the co-moving number density 
by integrating the LRG luminosity function. However, the different selection 
functions for low and high redshift LRGs (cut I and II, \citealt{eisenstein_annis_2001}) 
are difficult to model and could result in large uncertainties. Therefore, we
decide to limit the estimate to the volume-limited redshift range. 

The random catalogs contain 1000 times as many objects as the data sample.
The procedure follows the description in Section~\ref{randomcat}.


\section{AGN Samples}

In this paper we derive the AGN ACF for X-ray and optically-selected 
SDSS AGN samples. The selection of the different AGN samples is described below.

\subsection{RASS/SDSS AGN Samples}

The {\it ROSAT} All-Sky Survey (RASS, \citealt{voges_aschenbach_1999}) is 
currently the most sensitive all-sky X-ray survey. Although  
relatively shallow, this data set has a huge potential for science, especially 
when combined with other large-area spectroscopic surveys such as SDSS.  
\cite{anderson_voges_2003, anderson_margon_2007} positionally cross-correlated 
RASS sources with SDSS spectroscopic objects and classified RASS and SDSS-detected 
AGNs based on SDSS DR5.  They find 6224 AGNs with broad permitted emission lines in excess of 
of 1000 km\,s$^{-1}$ FWHM and 515 narrow permitted emission line AGNs 
matching RASS sources within 1 arcmin. More details on the sample selection 
are given in paper I Section~2.2 and \cite{anderson_voges_2003, anderson_margon_2007}.
The RASS/SDSS AGN sample is biased toward unabsorbed AGNs due to {\it ROSAT}'s 
soft energy band (0.1-2.4 keV). AGNs with broad emission lines and UV excess are,
in general, known to show little to no X-ray absorption and therefore account for the vast 
majority of the RASS/SDSS AGNs. 

\begin{figure}
  \centering
 \resizebox{\hsize}{!}{ 
  \includegraphics[bbllx=85,bblly=369,bburx=540,bbury=700]{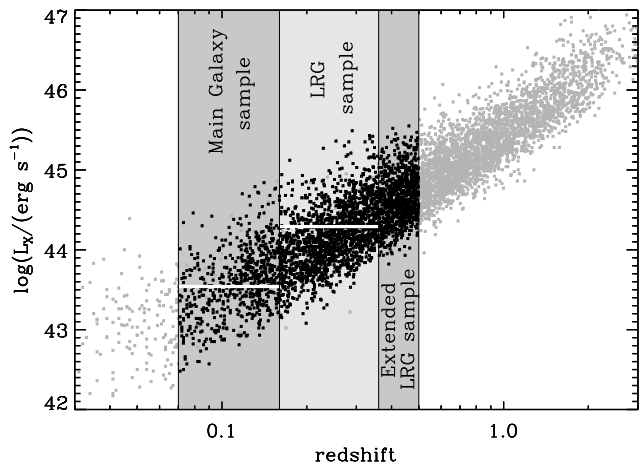}} 
      \caption{0.1--2.4 keV observed X-ray luminosity versus redshift for 
        the broad emission line AGN sample in SDSS DR5 determined by
               \cite{anderson_margon_2007}. Black symbols show objects 
               used here in the redshift range of $0.07 < z < 0.50$. The shaded 
               areas show the different AGN subsamples, labeled with the corresponding 
               tracer set for the cross-correlation measurement. 
               The horizontal white lines show the dividing line between 
               the lower X-ray luminosity and higher X-ray luminosity RASS/SDSS AGN subsamples 
               for $0.07 < z < 0.16$ log ($L_{\rm 0.1-2.4\,{\rm keV}}/{\rm [erg\,s^{-1}]})=43.54$
                and $0.16 < z < 0.36$ log ($L_{\rm 0.1-2.4\,{\rm keV}}/{\rm [erg\,s^{-1}]})=44.29$.}
         \label{LX_z}
\end{figure}

The latest available version of the RASS/SDSS AGN sample is based on DR5 (\citealt{anderson_margon_2007}).
Therefore, we reconfigure the sample to publically-available DR4+ geometry. We select broad emission line 
AGNs in different redshift ranges according to the redshift ranges of the available galaxy tracer sets 
(see Fig.~\ref{LX_z} and Table~\ref{samples}). 

To study the X-ray luminosity dependence of the clustering, we further subdivide the AGNs 
in the redshift ranges of $0.07<z<0.16$ and $0.16<z<0.36$  into relatively lower and higher X-ray luminosity 
samples. 
The RASS/SDSS AGNs in the redshift range of $0.36<z<0.50$ are not subdivided 
into a lower and higher X-ray luminosity sample, as the corresponding tracer set  
(extended LRGs) has a much lower number of objects than the other tracer sets. Consequently, 
there are too few pairs in the CCF and prohibitively large uncertainties to usefully compare 
the RASS/SDSS AGN ACF of lower and higher X-ray luminosity samples.   

We use the 0.1--2.4 keV observed luminosity reported by \cite{anderson_voges_2003, anderson_margon_2007}, which 
assumes a photon index of $\Gamma =2.5$ and is corrected for Galactic absorption. Our 0.1--2.4 keV luminosity 
cuts are log $(L_{\rm X}/[\rm{erg}\,\rm{s}^{-1}])=43.54$ for $0.07<z<0.16$ and 
log $(L_{\rm X}/[\rm{erg}\,\rm{s}^{-1}])=44.29$ for $0.16<z<0.36$ (see Fig.~\ref{LX_z}). 
Using $\Gamma =2.5$, the luminosity cuts correspond to 0.5--10 keV luminosities of 
$(L_{\rm X}/[10^{43}\,\rm{erg}\,\rm{s}^{-1}])= 1.5$ and 8.5, respectively.

A significant overlap between broad-line RASS/SDSS AGNs {\it and} the SDSS main galaxy sample tracer set 
exists only in the lowest redshift range ($0.07<z<0.16$). For the low luminosity broad-line AGNs in that 
redshift range, the AGN light does not dominate the 
optical spectrum. Hence, 186 objects of the total RASS/SDSS AGN sample of 629 objects 
are also classified as SDSS main galaxies (52 within the low $L_{\rm X}$ AGN sample, 
134 within the high $L_{\rm X}$ AGN sample). In the higher redshift samples, no match between RASS/SDSS 
AGN and the LRG tracer sets is found as the selection methods for LRGs and RASS/SDSS AGN 
are very different and high luminosity AGNs dominate the optical spectrum.
To quantify the effect of the overlap between both samples, we create a SDSS main galaxy sample 
that does not include objects that have also been classified as RASS/SDSS AGNs. Compared with the 
original CCF, the CCF of this new sample 
shows differences 
of less than 1\% in the pair counts on scales larger than 1.4 $h^{-1}$ Mpc. This is well below the 
statistical uncertainty of the CCF itself, therefore, the overlap between the AGN sample and the 
tracer set will not significantly affect our clustering measurements in the lowest redshift range.

We calculate the co-moving number densities given in Table~\ref{samples} as described in 
detail in paper I. For a given $R.A.$ and $Dec$ we compute the limiting observable RASS count 
rate and infer the absorption-corrected flux limit versus survey area for RASS/SDSS AGNs.
We then compute the co-moving volume available to each object ($V_{\rm a}$) for being 
included in the sample (\citealt{avni_bahcall_1980}). The co-moving number density follows by 
computing the sum of the available volume over each object $n_{\rm AGN}= \sum\limits_i 1/V_{{\rm a},i}$.

\subsubsection{Radio-quiet RASS/SDSS AGN Samples}
\label{radio_RASS}

Radio-loud AGNs are known to be more clustered than radio-quiet AGNs
and reside in very massive DMHs (e.g., \citealt{magliocchetti_maddox_2004}; 
\citealt{hickox_jones_2009}; \citealt{mandelbaum_li_2009}). Radio-loud AGNs are also more 
luminous in the X-rays than radio-quiet AGNs (\citealt{wilkes_tananbaum_1994}; 
\citealt{krumpe_lamer_2010b}). 

In paper I we find an X-ray luminosity dependence in the AGN clustering amplitude. 
One possible explanation is that the high $L_{\rm X}$ sample contains more radio-loud 
AGNs than the low $L_{\rm X}$ sample, and therefore the relative overabundance of radio-loud AGNs 
in the high $L_{\rm X}$ sample is causing the increase of the clustering amplitude.
To test this hypothesis, we construct radio-quiet RASS/SDSS AGN samples. 
\cite{anderson_voges_2003, anderson_margon_2007} list in their table of broad-line 
RASS/SDSS AGN if an object is also detected as a radio source. The radio information 
is taken from the Faint Images of the Radio Sky at Twenty centimeters (FIRST; 
\citealt{becker_white_1995}; \citealt{white_becker_1997}) using the NRAO Very Large Array. We 
therefore create new AGN subsamples by restricting all 
samples to the area covered by FIRST and excluding all FIRST detected sources, and 
refer to these subsamples as radio-quiet X-ray selected AGN samples. FIRST and SDSS 
cover almost the same area (FIRST has $\sim$7\% less area than SDSS DR4+ and 
$\sim$5\% less area than SDSS DR7). We also limit the tracer sets 
to the FIRST geometry when computing CCFs of radio-quiet AGN samples 
and the corresponding tracer set ACFs.

In Tables~\ref{samples}, \ref{table:acf_ccf}, 
and \ref{xagn_acf} we label these subsamples with the entry `(rq)'. 
Note that our approach is conservative, in that we do not apply the usual 
radio--to--optical flux density criteria of $R>10$ (\citealt{kellermann_sramek_1989}).
Instead we remove all radio-detected AGNs, which removes more objects than just those
that are technically defined as radio-loud. Our definition of radio-quiet 
is that the AGNs are not detected by FIRST. However,
for our goal of removing all radio-loud AGNs from the samples, the chosen 
procedure is adequate and the loss in a few additional AGNs will not 
significantly affect the 
clustering measurements and their uncertainties. 

In the redshift range of $0.07<z<0.16$, 127 out of 504 objects are classified as radio-quiet RASS/SDSS AGNs 
and SDSS main galaxies. Since we find no significant difference in the CCF of samples that include or exclude 
these objects, we use the full sample. No overlap is found between radio-quiet RASS/SDSS AGNs and tracer set 
objects at higher redshifts.

\begin{figure}
  \centering
 \resizebox{\hsize}{!}{ 
  \includegraphics[bbllx=75,bblly=369,bburx=544,bbury=700]{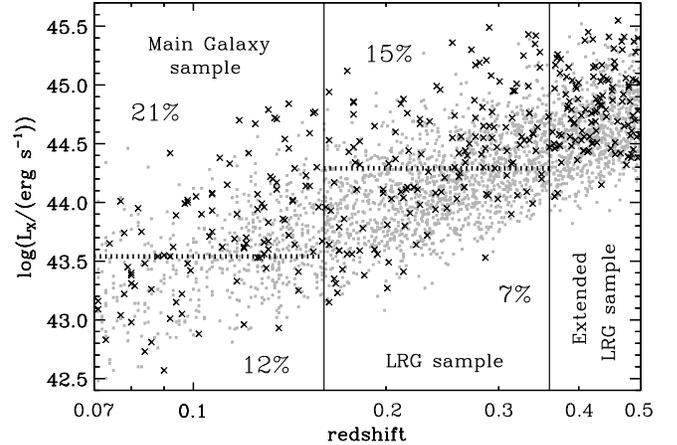}} 
      \caption{Location of FIRST radio-detected RASS/SDSS AGNs (crosses) in the 0.1-2.4 keV observed X-ray luminosity 
               versus redshift plane for the studied DR4+ X-ray samples. Gray dots show 
               RASS/SDSS AGNs without a FIRST radio detection, which we refer to as `radio-quiet' RASS/SDSS 
               AGN. We show only those objects that fall within the region covered by FIRST and SDSS. 
               The solid vertical lines indicate the redshift ranges of the different 
               samples, while the dotted horizontal lines show the luminosity 
               cuts used to create the lower and higher X-ray luminosity samples (see Fig.~\ref{LX_z}
               for details). The percentage values give the fraction of radio-detected RASS/SDSS AGNs 
               in the corresponding lower and higher $L_{\rm X}$ samples.}
         \label{LX_z_zoom_rq}
\end{figure}

In Fig.~\ref{LX_z_zoom_rq} we show the distribution of the RASS/SDSS AGNs which 
are flagged as radio sources by \cite{anderson_voges_2003, anderson_margon_2007},
which we remove to create the `radio-quiet' samples. 
Considering only objects that fall in regions covered by SDSS and FIRST, 17\% of 
all RASS/SDSS AGN have radio detections in the $0.07 <z<0.16$ range, 10\% at $0.16 <z<0.36$, 
and 14\% at $0.36 <z<0.50$. 
We also report the radio-detected AGN fraction in the corresponding 
lower and higher $L_{\rm X}$ samples as the percentage values listed in Fig.~\ref{LX_z_zoom_rq}.

\subsubsection{Narrow-line RASS/SDSS AGN Samples}
\label{narrowlineAGN}

\begin{figure}
  \centering
 \resizebox{\hsize}{!}{ 
  \includegraphics[bbllx=85,bblly=369,bburx=546,bbury=700]{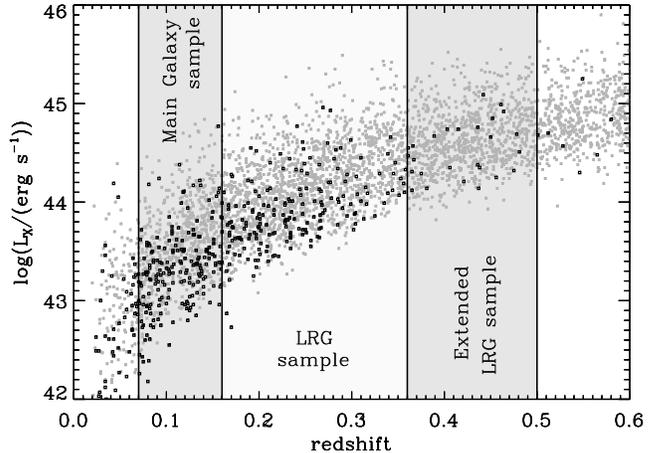}} 
      \caption{Location of narrow-line RASS/SDSS AGNs (black squares) in the 0.1-2.4 keV observed X-ray luminosity 
               versus redshift plane for the studied DR4+ X-ray samples. Gray dots show 
               broad-line RASS/SDSS AGNs. The filled areas illustrate the 
               redshift ranges of the different samples used here.}
         \label{LX_z_zoom_narrow}
\end{figure}

In order to test narrow-line versus broad-line AGN clustering, we
construct X-ray selected narrow-line AGN samples. 
In addition to the predominant broad-line AGNs, \cite{anderson_margon_2007} classified 
$\sim$7\% of all RASS/SDSS AGNs as X-ray emitting AGNs having narrower permitted 
emission lines. These 515 objects consist of X-ray emitting AGN subclasses such 
as narrow-line Seyfert 1's (NLS1s) (10\%), Seyfert 1.5 (29\%), Seyfert 1.8 (18\%), Seyfert 1.9 galaxies (22\%), 
and Seyfert 2 candidates (21\%). In total,  22\% of these objects also have a FIRST radio detection. 
NLS1s, Seyfert 1.5, 1.8, and 1.9 AGNs show a mix of narrow and broad permitted typical AGN line components, 
while Seyfert 2 candidates have only narrow permitted emission lines. The latter are called candidates 
because it is well known that a fair fraction of them turn out to be reclassified 
as NLS1, Seyfert 1.8, or Seyfert 1.9 when re-observed with significantly improved 
spectroscopy (e.g., \citealt{halpern_turner_1999}). Seyfert 1.5 and 1.8 galaxies have 
optical spectra with a broad line H$\beta$ component (exceeding FWHM values of 2500 km\,s$^{-1}$) 
at a very low flux level.
For more details about the different narrow-line AGN subclasses in this sample see \cite{anderson_voges_2003, anderson_margon_2007}.
Figure~\ref{LX_z_zoom_narrow} shows that the narrow-line RASS/SDSS AGNs are mainly identified only at 
lower redshifts, to $z\sim0.35$. Therefore, their classification as narrow-line AGNs is based 
on the permitted H$\beta$ and H$\alpha$ lines. 

The studied narrow-line RASS/SDSS AGNs are found to have, on average, lower observed 
X-ray luminosities than broad-line RASS/SDSS AGNs.
However, the vast majority of broad-line RASS/SDSS AGNs are known to be unabsorbed X-ray sources, therefore, their observed X-ray luminosity is equal to their intrinsic X-ray luminosity. 
On the other hand, narrow-line RASS/SDSS AGNs may be absorbed AGNs and their intrinsic 
X-ray luminosities could be higher than the observed ones given in 
Figure~\ref{LX_z_zoom_narrow}. Consequently, both classes of RASS/SDSS AGN samples may be very similar with respect to intrinsic X-ray luminosity.

Due to {\it ROSAT}'s soft energy range of 0.1--2.4 keV, 
narrow-line RASS/SDSS AGNs can be absorbed only by moderate column 
densities ($N_{\rm H}$). We simulate an X-ray spectrum 
with a photon index of $\Gamma=2.5$, galactic absorption of $N_{\rm H,gal}=2\times 10^{20}$\,cm$^{-2}$, 
and a typical redshift of $z=0.15$, in order  to estimate which intrinsic $N_{\rm H}$ values are detectable with 
{\it ROSAT}. Compared to an unabsorbed source, the flux in the 0.1--2.4 keV {\it ROSAT} band drops down to 61\% 
when an intrinsic column density of $N_{\rm H}= 10^{21}$\,cm$^{-2}$ is used 
(21\% for $N_{\rm H}= 10^{22}$\,cm$^{-2}$). We conclude that only objects with 
$N_{\rm H}\lesssim 10^{22}$\,cm$^{-2}$ are detected in the RASS. Most likely 
the narrow-line RASS/SDSS AGN sample consists mainly of a mixture of 
unabsorbed and only moderately absorbed (a few $N_{\rm H}= 10^{21}$\,cm$^{-2}$) AGNs.   

Although broad-line AGNs have a much lower projected space density than narrow-line AGNs and therefore 
yield, in general, more reliable identification with the RASS counterpart, 
\cite{anderson_voges_2003, anderson_margon_2007} successfully demonstrate that the 
narrow-line RASS/SDSS AGN are statistically very reliable identifications as well. They 
estimate that less than 5\% of the counterparts are spurious random chance positional 
coincidences. Convincing evidence derives from the distribution of positional offsets relative 
to the X-ray positional error and equal area annuli. Furthermore, the observed distribution of the ratios 
of RASS/SDSS X-ray--to--optical flux matches expectations for typical X-ray emitting AGNs.  
For details see \cite{anderson_voges_2003, anderson_margon_2007}. We conclude that the narrow-line
RASS/SDSS AGNs have a comparable high counterpart reliability as the broad-line RASS/SDSS AGNs.

In the redshift range of $0.07<z<0.16$ 31 out of 194 narrow-line RASS/SDSS AGNs are also classified as SDSS
main galaxies, while at $0.16<z<0.36$ only one object belongs to the narrow-line RASS/SDSS AGN 
sample and the LRG tracer set. As for the broad-line RASS/SDSS AGNs, we create a SDSS main galaxy sample 
that does not include galaxies that have been also classified as narrow-line RASS/SDSS AGNs and compute 
the corresponding CCF. On scales greater than 5 $h^{-1}$ Mpc, the pair counts differ by less than 0.1\% 
from the original CCF. Therefore, the overlap in the samples does not affect the clustering 
results for the narrow-line RASS/SDSS AGNs. 


\subsection{Optically-selected SDSS Samples}
\label{optically_selected_desc}

All of our optically-selected SDSS AGNs (called `quasars' in the SDSS 
literature) are drawn from \cite{schneider_richards_2010} and use the 
full SDSS survey (DR7).
The AGN candidate selection is described in detail in \cite{richards_fan_2002} and 
can be summarized as follows. The highest priority is given to FIRST-detected optical point sources. 
Then sources with non-stellar colors in the $ugriz$ photometry data are considered. 
Objects with photometric redshifts of $z\lesssim 3$ are targeted even if they 
are spatially resolved.  FIRST-detected optical point sources and $z\lesssim 3$ AGN candidates 
are required to have a Galactic extinction-corrected $i$ magnitude of 19.1. 
This selection method picks $\sim$ 18 objects per square degree which are 
followed up with the SDSS spectrograph. 
The resulting `primary' SDSS AGN sample is then supplemented by objects targeted by 
other SDSS spectroscopic selections (main galaxies, LRGs, RASS, stars, and serendipitous 
sources) that turned out to be AGNs. 
In the redshift range of interest ($0.07<z<0.50$), these secondary channels 
account for $\sim$9\% of the total SDSS AGN sample (vast majority from the galaxy 
target selection). If a significant number of SDSS AGNs from these secondary 
channels were originally selected from the LRG sample, which is known to be strongly 
clustered, it could bias the AGN clustering.
We verified that there is no overlap between the SDSS AGN and the LRG sample. Therefore,
we do not expect any influences on our AGN clustering measurements caused by the 
secondary selection channel.

\cite{schneider_richards_2010} (and references within) 
have constructed AGN catalogs based on different SDSS data releases.
They apply a luminosity selection of $M_i \le -22$ mag and require that objects have at least one 
emission line exceeding a FWHM of 1000 km\,s$^{-1}$. Objects that have a spectrum with only 
narrow permitted AGN-typical emission lines are removed. The absolute magnitude $M_i$ is 
computed by using the $i$ PSF Galactic extinction-corrected magnitude measurement and assuming
a typical AGN spectral energy slope.

Historically, quasars are defined as objects at the high end of the AGN luminosity function 
having $M_B\le -23$ mag (e.g., \citealt{schmidt_green_1983}). $M_i = -22$ mag corresponds to 
$M_B = -22.4$ mag for an typical AGN at $z=0$. The Schneider et al. SDSS AGN 
catalog papers use the $i$-band instead of the $B$-band in part because the $i$-filter is less affected 
by Galactic absorption. However, a significant disadvantage of this 
is that the host galaxy light may represent a larger contribution 
to the total flux then the AGN. AGNs near the $M_i = -22$ mag cut can be equally bright 
as the host galaxy, e.g., host galaxies at $z\sim 0.4$ with $i=19.1$ mag (within the detection limit of the 
SDSS AGN selection method) have $M_i = -22$ mag. Consequently, these AGNs may be less luminous than
their quoted optical magnitude. This effect is somewhat mitigated 
by the use of the PSF photometric data in the AGN selection but should be kept in mind when
interpreting our results.   

\begin{figure}
  \centering
 \resizebox{\hsize}{!}{ 
  \includegraphics[bbllx=64,bblly=369,bburx=546,bbury=700]{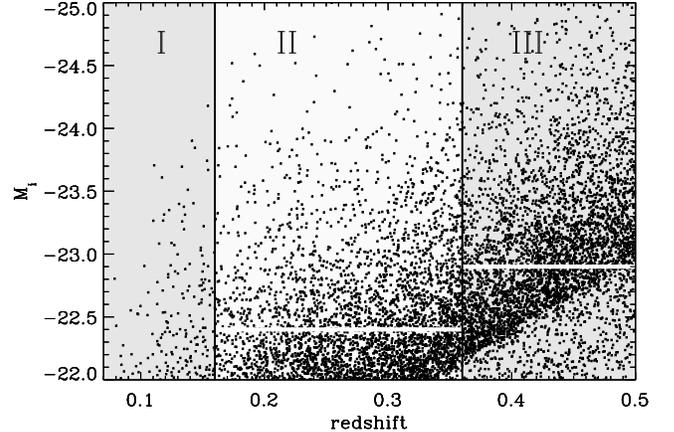}} 
      \caption{Absolute magnitude $M_i$ vs. redshift for optically-selected broad-line SDSS DR7 AGNs
               from \cite{schneider_richards_2010}. The vertical black lines 
               illustrate the redshift ranges of the different AGN samples and tracer sets, following
               Figs.~\ref{LX_z} and \ref{LX_z_zoom_narrow}. The white horizontal solid lines show the 
               cut used to create lower and higher $M_i$ optically-selected SDSS AGN subsamples.}
         \label{O_SDSS_overview}
\end{figure}

The AGN candidates selection by \cite{richards_fan_2002} has undergone constant 
modification to improve the efficiency. 
\cite{schneider_richards_2010, schneider_hall_2007}
use the final selection algorithm. AGNs in these catalogs have two spectroscopic 
target selection flags: BEST (final algorithm) and TARGET (actual algorithm during targeting).
BEST uses the latest photometric software and has the highest quality data. 
The continuous modification 
of the AGN selection method and the inclusion of AGNs not selected by the standard
selection means that the AGN catalogs are not statistically clean samples.  
The catalog of \cite{schneider_richards_2010} contains 105,783 spectroscopically 
confirmed AGNs which have luminosities of $M_i \le -22$, have at least one emission line 
exceeding a FWHM of 1000 km\,s$^{-1}$, have highly reliable redshifts, and are fainter than 
$i\sim 15$. We extract only objects that have a MODE flag of 'PRIMARY'. This procedure 
applies to 99\% of all objects and limits our sample to objects that have been spectroscopically 
followed up based on a target selection and are not blended. We use the BEST flags whenever 
TARGET and BEST are available.
We use optically-selected SDSS AGNs 
in the redshift ranges of the corresponding tracer sets 
($0.07<z<0.16$, $0.16<z<0.36$, $0.36<z<0.50$).
Except for the $0.07<z<0.16$ AGN sample, where the number of objects is very low, we 
further subdivide the samples into lower and higher $M_i$ subsamples (see Table~\ref{samples}). 
The $M_i$ cuts are chosen to yield approximately the same number of objects in the different 
luminosity subsamples. 
We use  $M_i=-22.4$ for the $0.16<z<0.36$ sample and $M_i=-22.9$ 
for the $0.36<z<0.50$ sample (Fig.~\ref{O_SDSS_overview}).

In the $0.07<z<0.16$ redshift range we find a high overlap between objects derived from the 
\cite{schneider_richards_2010} AGN sample and the SDSS main galaxy sample. 133 out of 177 
SDSS AGNs are also classified as SDSS main galaxies. As with the RASS/SDSS AGNs, we exclude 
these objects from the SDSS main galaxy sample and compute the CCF. The difference in the
CCF occurs only on the larger scales and is less than 1\%; therefore 
the overlap affects the clustering measurements very little.
At higher redshifts, there is no overlap between optically-selected SDSS AGN samples and SDSS 
LRGs used as our tracer sets. 
For the optically-selected SDSS AGNs we do not estimate the co-moving number densities in 
Table~\ref{samples}. Their selection function is very complex and requires a 
sophisticated modeling method, which is beyond the scope of this paper.

\subsubsection{Radio-quiet Optically-selected SDSS AGN Samples}
\label{radio_O_SDSS}
\begin{figure}
  \centering
 \resizebox{\hsize}{!}{ 
  \includegraphics[bbllx=111,bblly=367,bburx=516,bbury=722]{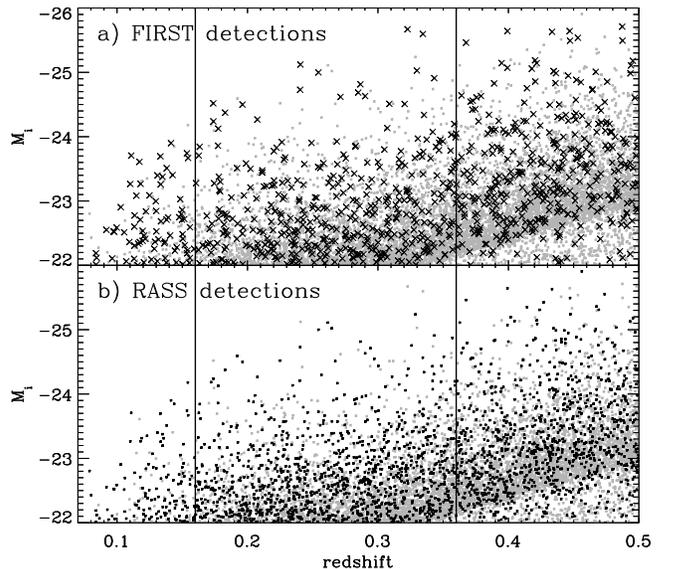}} 
      \caption{Similar to Figure~\ref{O_SDSS_overview}, here showing the location of 
                optically-selected broad-line FIRST-detected AGNs (top panel, crosses) and 
                RASS-detected AGNs (lower panel, asterisks) in the absolute $M_i$-magnitude--redshift  
               plane. In the top panel we select only objects that fall in regions covered by FIRST and SDSS. 
               The solid vertical lines show the different redshift ranges of the samples.}
         \label{O_SDSS_radio_X-ray}
\end{figure}

To test the impact of radio-loud AGNs on the clustering signal 
of optically-selected AGN samples, we further exclude FIRST radio-detected objects in the 
\cite{schneider_richards_2010} AGN samples. An AGN has a radio detection if its position 
coincidences with a FIRST catalog entry within 2 arcsec. We restrict all samples to the 
FIRST area to ensure that no 
radio-loud AGN is selected for the radio-quiet samples. Then we remove objects that have 
a measured FIRST radio flux in the AGN catalog. This is the same conservative 
approach of excluding all FIRST-detected AGNs as we do in Sect.~\ref{radio_RASS} for the 
radio-quiet RASS/SDSS AGN subsamples. Here we also use our definition of radio-quiet AGN 
samples for the optically selected AGNs, namely that radio--to--optical flux density $R$ 
based on the FIRST is $R=0$.
We also account for the restriction to the FIRST geometry in the corresponding tracer sets 
when we derive the ACF of the radio-quiet optically-selected AGN samples.
The resulting samples are labeled with the subsequent entry `(rq)' in Tables~\ref{samples}, 
\ref{table:acf_ccf}, and \ref{xagn_acf}.

Figure~\ref{O_SDSS_radio_X-ray} (top panel) shows the distribution of optically-selected FIRST-detected AGNs 
within the larger optically-selected AGN sample. A fraction of 44\% of the $0.07<z<0.16$ 
AGNs, 13\% of $0.16<z<0.36$ AGNs, and 9\% of the $0.36<z<0.50$ AGNs are detected by FIRST (considering 
only regions that are covered by FIRST and SDSS).
The fraction is a sensitive function of redshift, as it depends on luminosity. 
The higher $M_i$ subsamples contain more FIRST radio-detections  
(17\% in $0.16<z<0.36$ and 12\% in $0.36<z<0.50$) than 
the lower $M_i$ subsamples (8\% in $0.16<z<0.36$ and 5\% in $0.36<z<0.50$).
78 out of the 96 radio-quiet SDSS AGN in the redshift range $0.07<z<0.16$ are also classified 
as SDSS main galaxies.

\subsubsection{X-ray selected Optical AGN Samples}
The large number of X-ray detected optically-selected SDSS AGNs at redshifts above $z=0.16$
allows us to test how different selections of AGNs affect the clustering signal of 
broad-line AGNs. 
The catalog paper by \cite{schneider_richards_2010} lists for each individual 
AGN the relevant information if the object has a detection in the RASS faint or bright 
source catalog. Beside the construction of a radio-quiet optically-selected SDSS AGN sample 
(Section~\ref{radio_O_SDSS}), we create a sample that contains only 
optically-selected broad-line SDSS AGNs that are {\it not} detected in RASS. 
This sample has the subsequent entry `(noX)'
in the corresponding tables. Figure~\ref{O_SDSS_radio_X-ray} (panel b) shows the distribution of objects with RASS 
detections among the optically-selected SDSS AGN sample. A fraction of 70\% 
of the optically-selected SDSS AGNs are also detected by RASS in the redshift range of 
$0.07<z<0.16$. At higher redshifts RASS detects 32\% ($0.16<z<0.36$) and 
22\% ($0.36<z<0.50$) of the optically-selected broad-line SDSS AGNs. This is not surprising, as the low {\em ROSAT} 
sensitivity results in the RASS redshift distribution rising quickly to $z\sim 0.15$ and strongly
decreasing at higher redshift.

The optical and X-ray luminosities of broad emission line AGNs are known to 
be strongly correlated, the ratio of which is often expressed by the optical--to--X-ray 
spectral energy index $\alpha_{\rm ox}$ (e.g., \citealt{avni_tananbaum_1986}; 
\citealt{green_schartel_1995}; \citealt{steffen_strateva_2006}; \citealt{krumpe_lamer_2007}; 
\citealt{anderson_margon_2007}). Therefore, RASS detects 44\% of the AGNs in the higher $M_i$ sample 
at $0.16<z<0.36$ (29\% at $0.36<z<0.50$) but 
only 24\% of the AGNs in the lower $M_i$ sample at $0.16<z<0.36$ (17\% at $0.36<z<0.50$).

RASS contains only the highest flux X-ray emitting AGNs but has the 
complementary advantage of detecting lower AGN activity compared to the optically-selected 
broad-line SDSS AGNs. X-ray luminosities of log $(L_{\rm X}/[\rm{erg}\,\rm{s}^{-1}])> 42$ are 
a clear indicator of AGN activity, while in the optical 
a strong starlight component from the host galaxy can make it difficult to detect broad-line AGNs. 
The location of the RASS-detected AGNs in Fig.~\ref{O_SDSS_radio_X-ray} (panel b) 
indicates that RASS extends the detections of broad-line AGNs below the optical cut of  $M_i=-22$ mag 
used by \cite{schneider_richards_2010} at $z\lesssim 0.35$. 

Furthermore, we also create AGN samples that contain only optically-selected SDSS AGN that 
are also detected as RASS sources (Fig.~\ref{O_SDSS_radio_X-ray}, panel b). The samples are labeled with 
the subsequent entry `(onlyX)'. For these object we have both the X-ray and optical 
luminosities ($L_{\rm X}$, $M_i$).  
In general, as expected by the $\alpha_{\rm ox}$ connection, high/low $M_i$ corresponds to high/low 
$L_{\rm X}$. The fraction of FIRST-detected AGNs increases more with 
$M_i$ than with $L_{\rm X}$.
 
Finally, we design samples of optically-selected SDSS AGNs that are neither FIRST 
nor RASS detections. These samples only cover the FIRST area.
This selection results in 56\% of the $0.16<z<0.36$ and 67\% of the $0.36<z<0.50$ 
optically-selected SDSS AGNs.  We refer to these samples by the subsequent entry `(rq+noX)'.


\section{Clustering Analysis}
We measure the two-point correlation function $\xi(r)$ (\citealt{peebles_1980}), which measures 
the excess probability $dP$ above a Poisson distribution of finding an object in a volume 
element $dV$ at a distance $r$ from another randomly chosen object. The auto-correlation 
function (ACF) measures the spatial clustering of objects in the same sample, 
while the cross-correlation function (CCF) measures the clustering of objects in 
two different samples. We use the same approach as 
described in detail in paper I (Section~3). Here we reiterate the essential elements of our 
method.

We use the correlation estimator of \cite{davis_peebles_1983} in the form
\begin{equation}
\label{DD_DR}
 \xi(r)= \frac{DD(r)}{DR(r)} -1\ ,
\end{equation}
where $DD(r)$ is the number of data-data pairs with a separation $r$, and $DR(r)$ 
is the number data-random pairs; both pair counts have been normalized by the number density of
data and random points. 
We measure $\xi$ on a two-dimensional grid of separations $r_p$, perpendicular to the line of sight, 
and $\pi$, along the line of sight, to separate the effects of redshift space distortion due to 
peculiar velocities along the line of sight.
We obtain the  projected correlation function $w_p(r_p)$ by integrating $\xi(r_p,\pi )$ along
the $\pi$ direction.

As in paper I, we infer the AGN ACF from the CCF between AGNs and corresponding galaxy tracer set 
and the ACF of the tracer set using
\cite{coil_georgakakis_2009}: 
\begin{eqnarray}
\label{acf_rassagn}
 w_p(AGN|AGN) = \frac{\left[w_p(AGN|TRACE)\right]^2}{w_p(TRACE|TRACE)}\,,
\end{eqnarray}
where $w_p(AGN|AGN)$, $w_p(TRACE|TRACE)$ are the ACFs of the 
AGN and the corresponding tracer set, respectively, and $w_p(AGN|TRACE)$ is the 
CCF of the AGNs with the tracer set. In other words, we assume
that the CCF is the geometric mean of two ACFs, which has been verified 
to be valid by \cite{zehavi_zheng_2011} (Appendix A).

The CCF is computed by applying Eq.~\ref{DD_DR}
\begin{eqnarray}
\label{DD_DR2}
 \xi_{\rm AGN-TRACE} = \frac{D_{\rm AGN}\,D_{\rm TRACE}}{D_{\rm AGN}\,R_{\rm TRACE}}-1.
\end{eqnarray}
For our purposes, the use of this simple estimator has several major advantages and results in only 
a marginal loss in the signal-to-noise ratio when compared to more advanced estimators 
(e.g., \citealt{landy_szalay_1993}). The estimator in Eq.~\ref{DD_DR2} 
requires the generation of a random catalog only for the tracer set. The tracer sets have well-defined
selection functions and are, except for the extended LRG sample, 
volume-limited. Since the random catalog should exactly match all observational biases to 
minimize the systematic uncertainties, well understood selection effects are a key to 
generation proper random samples. The AGN samples suffer from very complex and hard to 
model selection functions. Therefore, a random catalog of X-ray selected RASS/SDSS AGN 
is subject to large systematic uncertainties due to the difficulty in accurately 
modeling the position-dependent sensitivity limit and the variation in the flux 
limit of the sources (caused by changing Galactic absorption over the sky 
and spectrum-dependent corrections). Optically-selected SDSS AGNs 
(see Section~\ref{optically_selected_desc}) rely on constantly modified 
selection algorithms and the acceptance of additional incomplete AGN selection 
methods. Consequently, the modeling of their selection function for the 
generation of a random catalog would be very challenging.

The errors in the adjacent bins in correlation measurements are not independent.
Poisson errors will significantly underestimate the uncertainties 
and should not be used for error calculations. Instead, we use
the jackknife resampling technique to estimate the measurement errors
as well as  the covariance matrix $M_{ij}$, which reflects the degree to which bin $i$ is 
correlated with bin $j$. The covariance matrix is used to obtain reliable 
power law fits to $w_p(r_p)$ by minimizing the correlated $\chi^2$ values. 
In our jackknife resampling, we divide the survey area into $N_{\rm T}=100$ subsection 
for the DR4+ geometry and 131 subsections for DR7, each of which is $\sim$50--60 deg$^2$.
These $N_{\rm T}$ jackknife-resampled correlation functions define the 
covariance matrix (Eq.~\ref{jackknife}):
\begin{eqnarray}
\label{jackknife}
 M_{ij} = \frac{N_{\rm T} -1}{N_{\rm T}} \left[\sum_{k=1}^{N_{\rm T}} \bigg(w_k(r_{p,i})-\langle w(r_{p,i})\rangle\bigg)\right.\nonumber\\
          \times \bigg(w_k(r_{p,j})-\langle w(r_{p,j})\rangle\bigg)\bigg] \,  
\end{eqnarray}
We calculate $w_p(r_p)$ $N_{\rm T}$ times, where each jackknife sample excludes one 
section and $w_k(r_{p,i})$ and $w_k(r_{p,j})$ are from the $k$-th jackknife samples of the 
AGN ACF and $\langle w(r_{p,i})\rangle$, $\langle w(r_{p,j})\rangle$ are the averages over all of the 
jackknife samples. The uncertainties represent a 1$\sigma$ (68.3\%) confidence interval. 

The generation of covariance matrix for the inferred AGN ACF considers 
the $N_{\rm T}$ jackknife-resampled correlation functions of the CCF (AGN and 
corresponding tracer set) and the tracer set ACF. For each jacknife sample 
we calculate the inferred AGN ACF by using Equation~\ref{acf_rassagn}. The resulting 
$N_{\rm T}$ $w_p(r_p)$ jackknife-resampled projected correlation functions of the inferred ACFs 
are then used to compute the covariance matrix of the inferred AGN ACF.

\subsection{Inferring the AGN Auto-correlation Function}
\label{inferringACF}
\begin{figure}
  \centering
 \resizebox{\hsize}{!}{ 
  \includegraphics[bbllx=85,bblly=365,bburx=361,bbury=718]{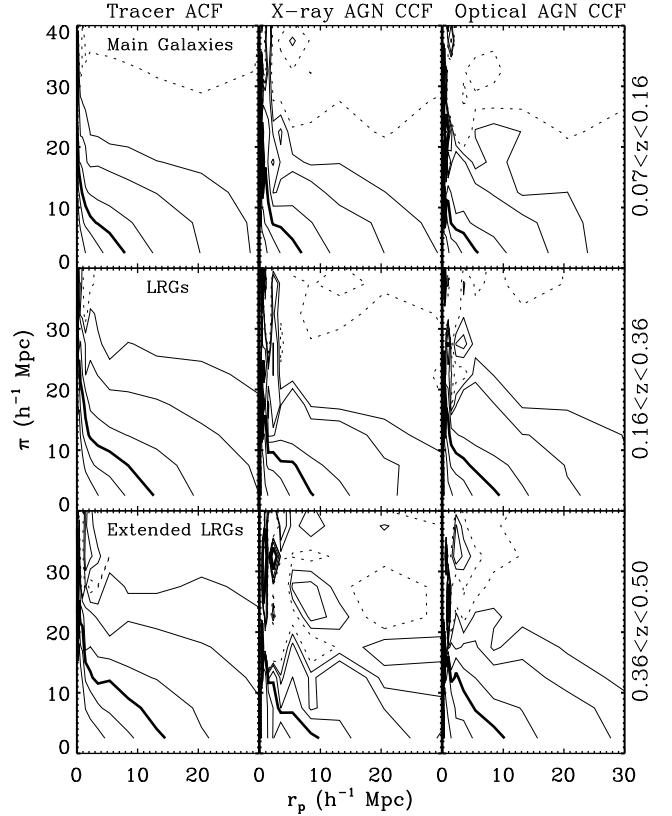}} 
      \caption{Contour plots of the auto-correlation functions $\xi(r_{\rm p},\pi)$ 
               of the different tracer sets
               (left panels, top -- SDSS main galaxies, middle -- LRGs, bottom -- 
               extended LRGs) and cross-correlation functions 
               of the X-ray selected (RASS/SDSS) AGN samples (middle panels) and the optically-selected 
               SDSS AGNs (right panels) with the corresponding tracer set. The shown CCFs use  
               in all cases the total AGN sample in the corresponding redshift range.
               In the $\pi$ direction we use a binning of 5 $h^{-1}$ Mpc. 
               Contour lines show constant correlation strength 
               for the two-dimensional correlation function $\xi(r_P,\pi )$. 
               The contour levels are 0.0 (dotted line), 0.1, 
               0.2, 0.5, 1.0 (thick solid line), 2.0, and 5.0.}
         \label{contour_acf}
\end{figure}

To infer the AGN ACF, we measure the CCF of 
the AGN sample with the tracer set and the ACF of the tracer set. 
In both cases we measure $r_p$ in a range of 0.05--40 $h^{-1}$ Mpc in 15 bins 
in a logarithmic scale. The upper 11 bins are identical with the bins used in paper I 
and cover the $r_p$ range of 0.3--40 $h^{-1}$ Mpc. Consequently, we extend the 
measurements by four additional bins to smaller scales. We compute $\pi$ in steps of 
5 $h^{-1}$ Mpc in a range of $\pi=0-200$ $h^{-1}$ Mpc. The resulting $\xi(r_P,\pi )$ are shown 
in Fig.~\ref{contour_acf} for the ACF of the tracer sets and 
for the CCF of the total AGN samples (X-ray \& optically-selected) 
with the corresponding tracer sets.

Although the projected correlation function is computed by integrating over $\pi$ to 
infinity (see paper I, Equations~5, 11), in practice an upper bound 
of the integration ($\pi_{\rm max}$) is used to include most of the correlated pairs, 
give stable solutions, and suppress the noise introduced by distant, uncorrelated pairs.
We compute $w_p(r_p)$ for a set of $\pi_{\rm max}$ ranging from 10--160 $h^{-1}$ Mpc 
in steps of 10 $h^{-1}$ Mpc. We then fit $w_p(r_p)$ over a $r_p$ range of 0.3--40 $h^{-1}$ Mpc 
with a fixed $\gamma = 1.9$ and determine the correlation length $r_{\rm 0}$ for the 
individual $\pi_{\rm max}$ measurements. As in paper I, we find that the LRG ACFs 
(LRG sample and extended LRG sample) saturate at $\pi_{\rm max}=80$ $h^{-1}$ Mpc. All CCFs 
and the main galaxy sample ACFs saturate at $\pi_{\rm max}=40$ $h^{-1}$ Mpc. 
In addition, above these values the corresponding correlation lengths 
do not change by more than 1$\sigma$, 
considering the increased uncertainties with increasing $\pi_{\rm max}$ values. 
The use of $\pi_{\rm max}=80$ $h^{-1}$ Mpc for the LRG ACFs and $\pi_{\rm max}=40$ $h^{-1}$ Mpc
for the main galaxy ACFs and all CCFs matches the $\pi_{\rm max}$ values used for 
these samples by other studies, e.g., \cite{zehavi_eisenstein_2005} (LRG) and 
\cite{zehavi_zheng_2005} (SDSS galaxies). The $w_p(r_p)$ CCFs for the different total AGN samples 
with the corresponding tracer sets are shown in Fig.~\ref{wp_CCF_AGN}, while the resulting 
power law fits for the ACFs and CCFs based on 
\begin{eqnarray}
 w_p(r_p) &=& r_p\left(\frac{r_{\rm 0}}{r_p}\right)^{\gamma}\,\frac{\Gamma(1/2)\Gamma((\gamma-1)/2)}{\Gamma(\gamma/2)},
\end{eqnarray}
where $\Gamma(x)$ is the Gamma function, are listed in Table~\ref{table:acf_ccf}.

\begin{deluxetable}{lccc}
\tabletypesize{\normalsize}
\tablecaption{Power Law Fits to the ACFs of the Tracer Sets and the CCFs of the AGN -- Tracer sets\label{table:acf_ccf}}
\tablewidth{0pt}
\tablehead{
\colhead{Sample} & \colhead{Redshift} &\colhead{$r_{\rm 0}$ }   & \colhead{$\gamma$}\\
\colhead{Name}   & \colhead{Range}    &\colhead{($h^{-1}$ Mpc)} & \colhead{}}
\startdata
\multicolumn{4}{c}{SDSS Tracer Sets}\\
main galaxy (DR4+)         & $0.07-0.16$ & 6.30$^{+0.12}_{-0.12}$ & 1.85$^{+0.02}_{-0.02}$ \\       
main galaxy (DR7)          & $0.07-0.16$ & 6.27$^{+0.12}_{-0.12}$ & 1.84$^{+0.02}_{-0.02}$ \\       
LRG (DR4+)                 & $0.16-0.36$ & 9.63$^{+0.14}_{-0.14}$ & 1.98$^{+0.02}_{-0.02}$ \\       
LRG (DR7)                  & $0.16-0.36$ & 9.54$^{+0.13}_{-0.13}$ & 1.95$^{+0.02}_{-0.02}$ \\       
extended LRG (DR4+)        & $0.36-0.50$ &10.90$^{+0.24}_{-0.24}$ & 1.91$^{+0.03}_{-0.03}$ \\       
extended LRG (DR7)         & $0.36-0.50$ &10.87$^{+0.19}_{-0.19}$ & 1.89$^{+0.03}_{-0.02}$ \\\hline 
\multicolumn{4}{c}{X-ray Selected AGN -- RASS/SDSS AGN}\\
total RASS-AGN             & $0.07-0.16$ & 5.79$^{+0.24}_{-0.25}$ & 1.84$^{+0.05}_{-0.05}$ \\       
total RASS-AGN(rq)         & $0.07-0.16$ & 5.78$^{+0.27}_{-0.27}$ & 1.86$^{+0.05}_{-0.05}$ \\       
low $L_{\rm X}$ RASS-AGN    & $0.07-0.16$ & 5.19$^{+0.34}_{-0.36}$ & 1.96$^{+0.10}_{-0.09}$ \\       
high $L_{\rm X}$ RASS-AGN   & $0.07-0.16$ & 6.07$^{+0.34}_{-0.35}$ & 1.79$^{+0.06}_{-0.06}$ \\       
low $L_{\rm X}$ RASS-AGN (rq)&$0.07-0.16$ & 5.49$^{+0.35}_{-0.36}$ & 1.94$^{+0.09}_{-0.09}$ \\       
high $L_{\rm X}$ RASS-AGN (rq)&$0.07-0.16$& 6.06$^{+0.42}_{-0.44}$ & 1.79$^{+0.06}_{-0.06}$ \\       
narrow line RASS-AGN       & $0.07-0.16$ & 4.99$^{+0.45}_{-0.48}$ & 1.82$^{+0.12}_{-0.10}$ \\       
                           &             &                      &                      \\
total RASS-AGN             & $0.16-0.36$ & 6.88$^{+0.27}_{-0.28}$ & 1.85$^{+0.04}_{-0.04}$ \\       
total RASS-AGN(rq)         & $0.16-0.36$ & 7.01$^{+0.28}_{-0.29}$ & 1.86$^{+0.05}_{-0.05}$ \\       
low $L_{\rm X}$ RASS-AGN    & $0.16-0.36$ & 6.22$^{+0.36}_{-0.38}$ & 1.85$^{+0.07}_{-0.06}$ \\       
high $L_{\rm X}$ RASS-AGN   & $0.16-0.36$ & 7.50$^{+0.38}_{-0.41}$ & 1.90$^{+0.07}_{-0.07}$ \\       
low $L_{\rm X}$ RASS-AGN (rq)& $0.16-0.36$ & 6.09$^{+0.37}_{-0.39}$& 1.85$^{+0.07}_{-0.06}$ \\       
high $L_{\rm X}$ RASS-AGN (rq)& $0.16-0.36$& 7.83$^{+0.39}_{-0.41}$& 1.94$^{+0.08}_{-0.08}$ \\      
narrow line RASS-AGN       & $0.16-0.36$ & 5.78$^{+0.74}_{-0.86}$ & 1.77$^{+0.14}_{-0.14}$ \\       
                           &             &                      &                      \\
total RASS-AGN             & $0.36-0.50$ & 6.82$^{+0.42}_{-0.44}$ & 1.97$^{+0.10}_{-0.09}$ \\       
total RASS-AGN(rq)         & $0.36-0.50$ & 6.67$^{+0.46}_{-0.49}$ & 2.08$^{+0.11}_{-0.11}$ \\\hline 
\multicolumn{4}{c}{Optically-selected AGN -- SDSS AGN}\\
total SDSS-AGN             & $0.07-0.16$ & 4.93$^{+0.41}_{-0.43}$ & 1.98$^{+0.12}_{-0.11}$ \\       
total SDSS-AGN(rq)         & $0.07-0.16$ & 4.70$^{+0.53}_{-0.58}$ & 2.11$^{+0.17}_{-0.14}$ \\       
                           &             &                      &                      \\
total SDSS-AGN             & $0.16-0.36$ & 6.91$^{+0.17}_{-0.18}$ & 1.91$^{+0.04}_{-0.04}$ \\       
total SDSS-AGN(rq)         & $0.16-0.36$ & 6.98$^{+0.20}_{-0.20}$ & 1.88$^{+0.04}_{-0.04}$ \\       
total SDSS-AGN (noX)       & $0.16-0.36$ & 6.93$^{+0.21}_{-0.22}$ & 1.93$^{+0.04}_{-0.04}$ \\
total SDSS-AGN (rq+noX)    & $0.16-0.36$ & 6.96$^{+0.23}_{-0.23}$ & 1.89$^{+0.05}_{-0.04}$ \\       
total SDSS-AGN (onlyX)     & $0.16-0.36$ & 7.10$^{+0.32}_{-0.33}$ & 1.85$^{+0.05}_{-0.05}$ \\       
low $M_i$ SDSS-AGN         & $0.16-0.36$ & 6.91$^{+0.24}_{-0.25}$ & 1.91$^{+0.05}_{-0.05}$ \\       
high $M_i$ SDSS-AGN       & $0.16-0.36$ & 6.74$^{+0.22}_{-0.22}$ & 1.90$^{+0.05}_{-0.05}$ \\       
low $M_i$ SDSS-AGN (rq)    & $0.16-0.36$ & 6.88$^{+0.26}_{-0.26}$ & 1.90$^{+0.05}_{-0.05}$ \\       
high $M_i$ SDSS-AGN (rq)   & $0.16-0.36$ & 6.91$^{+0.27}_{-0.28}$ & 1.87$^{+0.05}_{-0.05}$ \\       
                           &             &                      &                      \\
total SDSS-AGN             & $0.36-0.50$ & 7.21$^{+0.21}_{-0.22}$ & 1.87$^{+0.04}_{-0.04}$ \\       
total SDSS-AGN (rq)        & $0.36-0.50$ & 7.24$^{+0.22}_{-0.23}$ & 1.91$^{+0.04}_{-0.04}$ \\       
total SDSS-AGN (noX)       & $0.36-0.50$ & 7.02$^{+0.24}_{-0.25}$ & 1.81$^{+0.04}_{-0.04}$ \\       
total SDSS-AGN (rq+noX)    & $0.36-0.50$ & 7.08$^{+0.25}_{-0.26}$ & 1.84$^{+0.04}_{-0.04}$ \\       
low $M_i$ SDSS-AGN         & $0.36-0.50$ & 7.12$^{+0.30}_{-0.31}$ & 1.91$^{+0.06}_{-0.06}$ \\       
high $M_i$ SDSS-AGN        & $0.36-0.50$ & 6.96$^{+0.33}_{-0.36}$ & 1.81$^{+0.07}_{-0.07}$ \\       
low $M_i$ SDSS-AGN (rq)    & $0.36-0.50$ & 7.12$^{+0.32}_{-0.34}$ & 1.95$^{+0.06}_{-0.06}$ \\       
high $M_i$ SDSS-AGN (rq)   & $0.36-0.50$ & 7.16$^{+0.33}_{-0.36}$ & 1.84$^{+0.06}_{-0.06}$          
\enddata
\tablecomments{Values of $r_{\rm 0}$ and $\gamma$ obtained from a power law fit to 
  $w_{\rm p}(r_{\rm p})$ over the range $r_p$=0.3--40 
  $h^{-1}$ Mpc for all samples using the full covariance matrix. For the LRG and extended LRG ACFs,  
  we use  $\pi_{\rm max}=80$ $h^{-1}$ Mpc, while for all other ACFs and CCFs we use $\pi_{\rm max}=40$ $h^{-1}$ Mpc.
  See Table~\ref{samples} for the definition of the samples.}
\end{deluxetable}

Our values for the ACFs of the tracer sets agree well with measurements from other studies. 
Using a slightly different magnitude cut of $-22<M_r<-21$ in the redshift range $0.07<z<0.16$, 
\cite{zehavi_zheng_2005} find for the SDSS DR2 main galaxy sample a correlation length 
of $r_{\rm 0}=6.16\pm 0.17$ $h^{-1}$ Mpc
and a slope of $\gamma=1.85\pm0.03$, fitting over the range $r_p$=0.13--20 $h^{-1}$ Mpc.
\cite{zehavi_eisenstein_2005} study the clustering of $\sim$30,000 LRGs
with $-23.2<M_g^{0.3}<-21.2$  and measure $r_{\rm 0}=9.80\pm 0.20$ $h^{-1}$ Mpc and a slope 
of $\gamma=1.94\pm0.02$, fitting over the range $r_p$=0.3--30 $h^{-1}$ Mpc. Their clustering 
measurement in a sample of high luminosity LRGs ($-23.2<M_g^{0.3}<-21.8$, $0.16<z<0.44$) 
yields $r_{\rm 0}=11.21\pm 0.24$ $h^{-1}$ Mpc and $\gamma=1.92\pm0.03$.

In Table~\ref{xagn_acf} we list the redshift range, the median effective redshift of 
$N_{\rm CCF}(z)$ for the corresponding AGN samples, the derived best $r_{\rm 0}$ and $\gamma$ values 
based on power law fits, and $r_{\rm 0}$ for a power law fit with a fixed slope of $\gamma=1.9$.
The data are fitted over the range $r_p=0.3-15$ $h^{-1}$ Mpc to be consistent with paper I. 
Since we measure the CCF to infer the ACF, the resulting effective redshift 
distribution for the clustering signal is determined by both the redshift distribution 
of the tracer set and the AGN sample: $N_{\rm CCF}(z) = N_{\rm tracer}(z)*N_{\rm AGN}(z)$.

The clustering strength is commonly expressed in terms of the rms fluctuations within a sphere 
with a co-moving radius of 8 $h^{-1}$ Mpc ($\sigma_{\rm 8,AGN}$, see Equation~13 in paper I). We derive 
$\sigma_{\rm 8,AGN}$ from the best fit parameters of the power law fits. The uncertainties on $\sigma_{\rm 8,AGN}$ 
are derived from the $r_{\rm 0}$ versus $\gamma$ confidence contours of the one-parameter fit based
on a correlated $\chi^2= \chi^2_{\rm min} + 1.0$. Using $\sigma_{\rm 8,AGN}$, we further derive 
the bias parameter $b=\sigma_{\rm 8,AGN}(z)/\sigma_8(z)$ based on our power law fits and give these 
values in column `$b(z)$ PL-fits' of Table~\ref{xagn_acf}. The parameter $b$ indicates the clustering strength 
by comparing the observed AGN clustering to that of the underlying mass distribution from linear growth theory 
(\citealt{hamilton_2001}), with $\sigma_8(z)=D(z) \sigma_8(z=0)$, where $D(z)=D_1(z)/D_1(z=0)$ is the linear growth factor 
(Section 7.5 in \citealt{dodelson_2003}).
We use $\sigma_8(z=0)= 0.8$ (see Section~\ref{introduction}).  
The bias quantifies the amplification factor of the contrast of the object distribution
with respect to that of the dark matter density distribution. 
The uncertainties of $b$ are derived from the standard deviation of $\sigma_{\rm 8,AGN}$. 

Column `$b(z)$ HOD' of Table~\ref{xagn_acf} lists the bias parameter derived using halo occupation 
distribution (HOD) modeling, which is described below in Section~\ref{hod_model}.
 Using Equation~8 of \cite{sheth_mo_2001} and the improved fit  for this equation given by 
\cite{tinker_weinberg_2005}, we compute the expected large-scale Eulerian bias factor for different 
DMH masses at different redshifts. Comparing the observed $b$ value from HOD modeling
with the DMH bias factor from $\Lambda$CDM cosmological simulations provides an estimate 
of the typical DMH mass ($b_{\rm DMH}(M_{\rm DMH}^{\rm typ}) = b_{\rm OBS,HOD}$) in which the 
different AGN samples reside, listed in the  last column of Table~\ref{xagn_acf}.

\begin{deluxetable*}{lcccccccc}
\tabletypesize{\normalsize}
\tablecaption{Power Law Fits to the Inferred AGN ACF and Derived Quantities.\label{xagn_acf}}
\tablewidth{0pt}
\tablehead{
\colhead{Sample} & \colhead{Redshift} &\colhead{Median} & \colhead{$r_{\rm 0}$}   & \colhead{$\gamma$}         &\colhead{$r_{\rm 0,\gamma =1.9}$}    &\colhead{$b(z)$}  &\colhead{$b(z)$}     &\colhead{log $M_{\rm DMH}^{\rm typ}$}\\
\colhead{Name} & \colhead{Range} & \colhead{$z_{\rm eff}$} & \colhead{($h^{-1}$ Mpc)} & \colhead{} &\colhead{($h^{-1}$ Mpc)       } &\colhead{PL-fits} &\colhead{HOD}        &\colhead{($h^{-1}$ $M_{\odot}$)}}
\startdata
\multicolumn{9}{c}{X-ray Selected AGN -- RASS/SDSS AGN}\\
total RASS-AGN                  & $0.07-0.16$ & 0.13 &4.96$^{+0.41}_{-0.44}$ & 1.80$^{+0.10}_{-0.10}$& 4.84$^{+0.38}_{-0.41}$ &1.19$^{+0.08}_{-0.09}$ &1.23$^{+0.09}_{-0.08}$ &13.22$^{+0.13}_{-0.12}$\\
total RASS-AGN(rq)              & $0.07-0.16$ & 0.14 &4.42$^{+0.69}_{-0.77}$ & 1.85$^{+0.17}_{-0.14}$& 4.33$^{+0.62}_{-0.71}$ &1.08$^{+0.15}_{-0.17}$ &1.25$^{+0.11}_{-0.08}$ &13.24$^{+0.14}_{-0.12}$\\
low $L_{\rm X}$ RASS-AGN         & $0.07-0.16$ & 0.13 &3.84$^{+0.52}_{-0.59}$ & 2.10$^{+0.23}_{-0.20}$& 3.85$^{+0.55}_{-0.63}$ &0.98$^{+0.13}_{-0.15}$ &1.11$^{+0.11}_{-0.05}$ &13.03$^{+0.18}_{-0.09}$\\
high $L_{\rm X}$ RASS-AGN        & $0.07-0.16$ & 0.14 &5.35$^{+0.65}_{-0.74}$ & 1.66$^{+0.11}_{-0.12}$& 4.94$^{+0.56}_{-0.63}$ &1.25$^{+0.12}_{-0.15}$ &1.33$^{+0.16}_{-0.14}$ &13.35$^{+0.17}_{-0.20}$\\
low $L_{\rm X}$ RASS-AGN (rq)    & $0.07-0.16$ & 0.12 &4.18$^{+0.77}_{-0.91}$ & 2.11$^{+0.30}_{-0.23}$& 4.28$^{+0.79}_{-0.95}$ &1.07$^{+0.19}_{-0.21}$ &1.17$^{+0.11}_{-0.08}$ &13.15$^{+0.15}_{-0.14}$\\
high $L_{\rm X}$ RASS-AGN (rq)   & $0.07-0.16$ & 0.14 &4.67$^{+1.07}_{-1.34}$ & 1.61$^{+0.18}_{-0.17}$& 3.86$^{+0.89}_{-1.13}$ &1.10$^{+0.22}_{-0.26}$ &1.36$^{+0.24}_{-0.18}$ &13.38$^{+0.25}_{-0.24}$\\
narrow line RASS-AGN            & $0.07-0.16$ & 0.13 &3.14$^{+0.81}_{-1.23}$ & 1.64$^{+0.41}_{-0.28}$& 3.10$^{+0.70}_{-0.88}$ &0.80$^{+0.16}_{-0.23}$ &1.24$^{+0.18}_{-0.19}$ &13.24$^{+0.22}_{-0.32}$\\
                                &             &                        &                     &                      &                      &                     & \\
total RASS-AGN                  & $0.16-0.36$ & 0.27 &4.05$^{+0.43}_{-0.52}$ & 1.63$^{+0.12}_{-0.12}$& 4.07$^{+0.35}_{-0.38}$ &1.06$^{+0.09}_{-0.11}$ &1.30$^{+0.10}_{-0.08}$ &13.17$^{+0.12}_{-0.11}$\\
total RASS-AGN(rq)              & $0.16-0.36$ & 0.26 &4.50$^{+0.40}_{-0.46}$ & 1.73$^{+0.12}_{-0.12}$& 4.50$^{+0.36}_{-0.39}$ &1.16$^{+0.09}_{-0.11}$ &1.27$^{+0.08}_{-0.08}$ &13.14$^{+0.11}_{-0.12}$\\
low $L_{\rm X}$ RASS-AGN         & $0.16-0.36$ & 0.24 &3.16$^{+0.51}_{-0.70}$ & 1.78$^{+0.43}_{-0.28}$& 3.18$^{+0.42}_{-0.48}$ &0.85$^{+0.11}_{-0.13}$ &1.17$^{+0.14}_{-0.12}$ &13.01$^{+0.21}_{-0.22}$\\
high $L_{\rm X}$ RASS-AGN        & $0.16-0.36$ & 0.31 &5.06$^{+0.65}_{-0.87}$ & 1.82$^{+0.17}_{-0.18}$& 5.19$^{+0.58}_{-0.65}$ &1.34$^{+0.19}_{-0.22}$ &1.48$^{+0.12}_{-0.15}$ &13.34$^{+0.12}_{-0.17}$\\
low $L_{\rm X}$ RASS-AGN (rq)    & $0.16-0.36$ & 0.24 &3.62$^{+0.63}_{-0.72}$ & 1.85$^{+0.37}_{-0.26}$& 3.59$^{+0.59}_{-0.70}$ &0.95$^{+0.14}_{-0.17}$ &1.18$^{+0.14}_{-0.13}$ &13.03$^{+0.20}_{-0.24}$\\
high $L_{\rm X}$ RASS-AGN (rq)   & $0.16-0.36$ & 0.30 &6.00$^{+0.65}_{-0.77}$ & 1.97$^{+0.18}_{-0.18}$& 5.90$^{+0.66}_{-0.74}$ &1.63$^{+0.22}_{-0.25}$ &1.50$^{+0.20}_{-0.13}$ &13.37$^{+0.18}_{-0.14}$\\
                                &             &                        &                     &                      &                      &                     & \\
total RASS-AGN                  & $0.36-0.50$ & 0.42 &3.24$^{+0.96}_{-2.13}$ & 1.59$^{+0.40}_{-0.41}$& 3.83$^{+0.55}_{-0.63}$ &0.96$^{+0.22}_{-0.54}$ &1.02$^{+0.14}_{-0.09}$ &12.51$^{+0.28}_{-0.25}$\\  
total RASS-AGN(rq)              & $0.36-0.50$ & 0.42 &4.13$^{+0.81}_{-1.94}$ & 1.98$^{+0.59}_{-0.56}$& 4.01$^{+0.80}_{-0.98}$ &1.20$^{+0.34}_{-0.47}$ &0.99$^{+0.11}_{-0.06}$ &12.43$^{+0.25}_{-0.17}$\\\hline
\multicolumn{9}{c}{Optically-Selected AGN -- SDSS AGN}\\
total SDSS-AGN                  & $0.07-0.16$ & 0.14 &3.90$^{+0.62}_{-0.70}$ & 1.99$^{+0.29}_{-0.25}$& 3.94$^{+0.62}_{-0.72}$ &0.98$^{+0.15}_{-0.17}$ &0.95$^{+0.17}_{-0.10}$ &12.67$^{+0.37}_{-0.31}$\\
total SDSS-AGN(rq)              & $0.07-0.16$ & 0.15 &4.16$^{+0.85}_{-1.07}$ & 2.12$^{+0.55}_{-0.39}$& 4.26$^{+0.83}_{-1.02}$ &1.08$^{+0.20}_{-0.23}$ &0.94$^{+0.24}_{-0.13}$ &12.63$^{+0.50}_{-0.45}$\\
                                &             &                        &                     &                      &                      &                     & \\
total SDSS-AGN                  & $0.16-0.36$ & 0.31 &4.80$^{+0.24}_{-0.27}$ & 1.79$^{+0.09}_{-0.09}$& 4.81$^{+0.23}_{-0.24}$ &1.26$^{+0.07}_{-0.07}$ &1.29$^{+0.05}_{-0.05}$ &13.11$^{+0.07}_{-0.07}$\\
total SDSS-AGN(rq)              & $0.16-0.36$ & 0.32 &4.78$^{+0.29}_{-0.33}$ & 1.72$^{+0.10}_{-0.10}$& 4.77$^{+0.26}_{-0.27}$ &1.25$^{+0.07}_{-0.07}$ &1.31$^{+0.08}_{-0.06}$ &13.13$^{+0.10}_{-0.09}$\\
total SDSS-AGN (noX)            & $0.16-0.36$ & 0.32 &4.95$^{+0.31}_{-0.33}$ & 1.85$^{+0.10}_{-0.10}$& 4.94$^{+0.30}_{-0.32}$ &1.31$^{+0.09}_{-0.09}$ &1.26$^{+0.08}_{-0.06}$ &13.06$^{+0.11}_{-0.09}$\\
total SDSS-AGN (rq+noX)         & $0.16-0.36$ & 0.32 &4.80$^{+0.33}_{-0.35}$ & 1.78$^{+0.12}_{-0.11}$& 4.72$^{+0.31}_{-0.33}$ &1.27$^{+0.07}_{-0.09}$ &1.29$^{+0.10}_{-0.07}$ &13.10$^{+0.13}_{-0.10}$\\
total SDSS-AGN (onlyX)          & $0.16-0.36$ & 0.29 &4.70$^{+0.50}_{-0.60}$ & 1.71$^{+0.12}_{-0.12}$& 4.77$^{+0.43}_{-0.47}$ &1.22$^{+0.12}_{-0.15}$ &1.35$^{+0.10}_{-0.11}$ &13.21$^{+0.12}_{-0.15}$\\
low $M_i$ SDSS-AGN              & $0.16-0.36$ & 0.31 &4.73$^{+0.36}_{-0.43}$ & 1.72$^{+0.14}_{-0.13}$& 4.73$^{+0.32}_{-0.34}$ &1.23$^{+0.09}_{-0.10}$ &1.31$^{+0.11}_{-0.09}$ &13.14$^{+0.13}_{-0.13}$\\
high $M_i$ SDSS-AGN             & $0.16-0.36$ & 0.32 &4.36$^{+0.35}_{-0.43}$ & 1.75$^{+0.12}_{-0.12}$& 4.48$^{+0.30}_{-0.32}$ &1.17$^{+0.08}_{-0.11}$ &1.24$^{+0.08}_{-0.08}$ &13.03$^{+0.11}_{-0.13}$\\
low $M_i$ SDSS-AGN (rq)         & $0.16-0.36$ & 0.31 &4.39$^{+0.44}_{-0.60}$ & 1.65$^{+0.16}_{-0.15}$& 4.50$^{+0.34}_{-0.37}$ &1.16$^{+0.10}_{-0.13}$ &1.30$^{+0.09}_{-0.12}$ &13.13$^{+0.11}_{-0.18}$\\
high $M_i$ SDSS-AGN (rq)        & $0.16-0.36$ & 0.32 &4.53$^{+0.44}_{-0.56}$ & 1.70$^{+0.14}_{-0.13}$& 4.64$^{+0.37}_{-0.40}$ &1.19$^{+0.10}_{-0.13}$ &1.30$^{+0.12}_{-0.08}$ &13.12$^{+0.14}_{-0.12}$\\
                                &             &                        &                     &                      &                      &                     & \\
total SDSS-AGN                  & $0.36-0.50$ & 0.42  &4.41$^{+0.36}_{-0.44}$ & 1.69$^{+0.10}_{-0.10}$& 4.58$^{+0.29}_{-0.31}$ &1.23$^{+0.09}_{-0.11}$ &1.33$^{+0.07}_{-0.08}$ &13.05$^{+0.09}_{-0.12}$\\
total SDSS-AGN (rq)             & $0.36-0.50$ & 0.42  &4.66$^{+0.36}_{-0.43}$ & 1.78$^{+0.11}_{-0.11}$& 4.76$^{+0.32}_{-0.34}$ &1.30$^{+0.09}_{-0.12}$ &1.32$^{+0.08}_{-0.08}$ &13.03$^{+0.11}_{-0.11}$\\
total SDSS-AGN (noX)            & $0.36-0.50$ & 0.42  &2.93$^{+0.51}_{-0.66}$ & 1.39$^{+0.10}_{-0.10}$& 3.16$^{+0.28}_{-0.30}$ &0.92$^{+0.09}_{-0.12}$ &1.42$^{+0.08}_{-0.09}$ &13.16$^{+0.09}_{-0.11}$\\
total SDSS-AGN (rq+noX)         & $0.36-0.50$ & 0.42  &3.79$^{+0.46}_{-0.58}$ & 1.52$^{+0.10}_{-0.10}$& 3.80$^{+0.32}_{-0.35}$ &1.09$^{+0.09}_{-0.11}$ &1.43$^{+0.10}_{-0.09}$ &13.17$^{+0.11}_{-0.11}$\\
low $M_i$ SDSS-AGN              & $0.36-0.50$ & 0.40  &4.59$^{+0.47}_{-0.61}$ & 1.72$^{+0.16}_{-0.16}$& 4.70$^{+0.40}_{-0.43}$ &1.26$^{+0.12}_{-0.15}$ &1.29$^{+0.11}_{-0.09}$ &13.02$^{+0.14}_{-0.14}$\\
high $M_i$ SDSS-AGN             & $0.36-0.50$ & 0.43  &2.92$^{+0.66}_{-1.02}$ & 1.48$^{+0.16}_{-0.18}$& 3.42$^{+0.38}_{-0.42}$ &0.90$^{+0.14}_{-0.19}$ &1.27$^{+0.11}_{-0.10}$ &12.95$^{+0.15}_{-0.15}$\\
low $M_i$ SDSS-AGN (rq)         & $0.36-0.50$ & 0.40  &4.68$^{+0.48}_{-0.61}$ & 1.82$^{+0.18}_{-0.17}$& 4.74$^{+0.45}_{-0.49}$ &1.31$^{+0.14}_{-0.17}$ &1.26$^{+0.13}_{-0.12}$ &12.97$^{+0.18}_{-0.19}$\\
high $M_i$ SDSS-AGN (rq)        & $0.36-0.50$ & 0.44  &3.59$^{+0.61}_{-0.92}$ & 1.56$^{+0.16}_{-0.17}$& 3.87$^{+0.42}_{-0.47}$ &1.05$^{+0.14}_{-0.19}$ &1.35$^{+0.12}_{-0.10}$ &13.05$^{+0.14}_{-0.14}$ 
\tablecomments{Values of $r_{\rm 0}$, $\gamma$, and $r_{\rm 0,\gamma =1.9}$ are obtained from a  fit to $w_{\rm p}(r_{\rm p})$ over the range $r_p=$ 0.3--15 $h^{-1}$ Mpc for 
               all samples using the full error covariance matrix and minimizing the correlated $\chi^2$ values. 
               The given bias parameters, $b=\sigma_{\rm 8,AGN}(z)/\sigma_8(z)$, are based on the best power law fit parameter and 
               from HOD modeling (HOD). To derive log $M_{\rm DMH}^{\rm typ}$, we use the bias parameter from HOD modeling.
               See Table~\ref{samples} for the definition of the samples.}
\end{deluxetable*}


\subsection{Robustness of the Clustering Measurements}
In this section, we verify the stability of our result against possible observational biases and 
systematic effects.  As shown in paper I,  the somewhat non-contiguous coverage of the SDSS DR4+ 
survey does not influence the clustering results significantly, given the uncertainties. For DR7 the 
situations improves as the SDSS geometry is much more contiguous. 
We verify that the number of random points is large enough to
lead to a high number of pair counts at the smallest scales measured and 
not introduce noise.

\cite{zehavi_zheng_2005} note that the largest structure detected in 
SDSS (the Sloan Great Wall) influences the galaxy clustering significantly
for samples with $M^{0.1}_r <-21$ and $z<0.1$.
Therefore, we explore this effect on our SDSS main galaxy 
sample. We measure the ACF for a main galaxy sample that excludes the 
Sloan Great Wall ($165<RA<210$ and $-5<Dec<5$). 
We find a correlation length of $r_{\rm 0}=6.35\pm 0.12$ $h^{-1}$ Mpc and 
$\gamma=1.83\pm0.02$, which agrees well with the clustering measurements 
in which the Sloan Great Wall is included (see Table~\ref{table:acf_ccf}). 
We conclude that our main galaxy ACFs and the CCFs using the main galaxy sample as a tracer 
set are not affected by this supercluster at $z\sim0.08$. 

For several of the AGN subsamples split by luminosity, we have tested that
 slightly changing the 
luminosity cuts by up to $\pm0.2$ mags (both brighter and fainter) 
does not significantly change the measured CCFs. 
The combination of the different tests listed above provides convincing evidence that our 
results are not significantly influenced by systematic effects and demonstrates their robustness.


\section{Bias from the HOD modeling}
\label{hod_model}

In paper II, we develop a novel method to infer the halo occupation distribution
(HOD) of RASS/SDSS AGNs directly from the well-constrained cross-correlation function 
of RASS/SDSS AGNs with LRGs. 
The results from paper II show that the linear bias parameters and typical DMH masses 
derived from the best power law fits down to $r_{\rm p}\approx 0.3$ $h^{-1}$ Mpc are subject 
to systematic errors. This is mainly because the power law fits include scales in the non-linear 
regime ($r_{\rm p}\la 1.5$ $h^{-1}$ Mpc), where the contribution from pairs of objects that 
belong to the same DMH (the one halo term) is substantial. In this non-linear regime the bias-DMH mass 
relation based on linear theory, in principle, 
should not be applied. Another, less significant source of systematic error is that even 
in the linear regime ($r_{\rm p}\ga 1.5$ $h^{-1}$ Mpc), the underlying matter correlation function 
deviates from a power law. In order to avoid these issues, \cite{allevato_2011} derive the 
bias parameters of their AGN samples in the XMM-COSMOS survey by modeling their AGN ACFs 
at $r_{\rm p}\ga 1.5$ $h^{-1}$ Mpc with $b^2_{\rm AGN}\xi_{\rm DM}^{2-h}$, where $b_{\rm AGN}$ is 
the AGN bias parameter and $\xi_{\rm DM}^{2-h}$ is the two halo term of the dark matter 
correlation function modeled as the Fourier Transform of the linear power spectrum.

In this paper, instead, we use the HOD modeling developed in paper II to derive the
bias parameter down to  $r_{\rm p}\ga 0.7$  $h^{-1}$ Mpc, instead of limiting ourselves to 
$r_{\rm p}\ga 1.5$ $h^{-1}$ Mpc. This allows for better constraints on $b_{\rm A}$, especially in 
cases where the two halo dominated (linear) regime extends below $r_{\rm p}\approx 1.5$ $h^{-1}$ Mpc, 
and a better treatment of the cases where the one halo term contribution is still important at 
$r_{\rm p}\ga 1.5$ $h^{-1}$ Mpc. Thus, in our approach, the main constraints are derived from
the two halo term, while including the one halo term contribution in the model serves as a first-order
perturbation from linear theory.

Paper II discusses the HOD modeling of the CCF between RASS/SDSS AGNs and LRGs at 
$0.16<z<0.36$ for three different models: in model A all AGNs are satellites within the same DMH 
as the LRGs, while models B and C include different realizations of the cases where central and 
satellite AGNs are included and explicitly parameterized. 
See paper II for details of these models.
We repeat this exercise here for the AGN samples used in this paper to derive their bias parameters. 
The detailed results of the extensive HOD modeling will be presented in a separate 
paper (Miyaji et al. in preparation), where a number of new improvements
in the modeling over that presented in paper II will be included. 
In this paper, we follow exactly the method presented in paper II. Here we reiterate the 
main procedure:
\begin{enumerate} 
\item Determine the central and satellite HODs of the tracer set galaxies ($\langle
N_{\rm G,c} \rangle (M_{\rm h})$ and $\langle N_{\rm G,s} \rangle (M_{\rm h})$, respectively) 
from their ACF. The space density constraint is additionally used for volume-limited samples (the 
SDSS main galaxy and the LRG sample).   
\item Using the derived tracer set HODs ($\langle N_{\rm G,c} \rangle (M_{\rm h})$ \& 
$\langle N_{\rm G,s} \rangle (M_{\rm h})$) and using a parameterized model
of the AGN central and satellite HODs ($\langle N_{\rm A,c} \rangle (M_{\rm h})$ 
\& $\langle N_{\rm A,s} \rangle (M_{\rm h})$), we fit the measured CCF between the 
AGN sample and the tracer set to constrain the AGN HODs. Since the galaxy ACFs have much
higher statistical accuracy, the uncertainties in the tracer set HODs are negligible compared 
to the AGN HOD constraints.
\item  Derive the bias parameter of the AGN sample using:
  \begin{equation}
  b_{\rm A}= \frac{\int b_{\rm h}(M_{\rm h}) \langle N_{\rm A} \rangle (M_{\rm h})
\phi(M_{\rm h}) dM_{\rm h}}{\int \langle N_{\rm A}\rangle (M_{\rm h}) \phi(M_{\rm
h})dM_{\rm h}},
  \label{eq:b_agn}
  \end{equation}
where $\langle N_{\rm A}\rangle (M_{\rm h})=\langle N_{\rm A,c}\rangle (M_{\rm
h})+\langle N_{\rm A,s}\rangle (M_{\rm h})$, $b_{\rm h}(M_{\rm h})$ is the bias of 
DMHs with a mass $M_{\rm h}$, and $\phi(M_{\rm h})$ is the DMH mass function.
We use Equation~8 of \citet{sheth_mo_2001} with parameters from 
\citet{tinker_weinberg_2005} for $b_{\rm h}(M_{\rm h})$. The 1$\sigma$ uncertainties of 
$b_{\rm A}$ corresponds to the $\Delta \chi^2 \le 1$ region in the parameter space 
of the $\langle N_{\rm A} \rangle (M_{\rm h})$ model.  
\end{enumerate}

\subsection{The HODs of the tracer sets}

\begin{figure*}
  \centering
 \resizebox{\hsize}{!}{ 
  \includegraphics[bbllx=32,bblly=172,bburx=551,bbury=670]{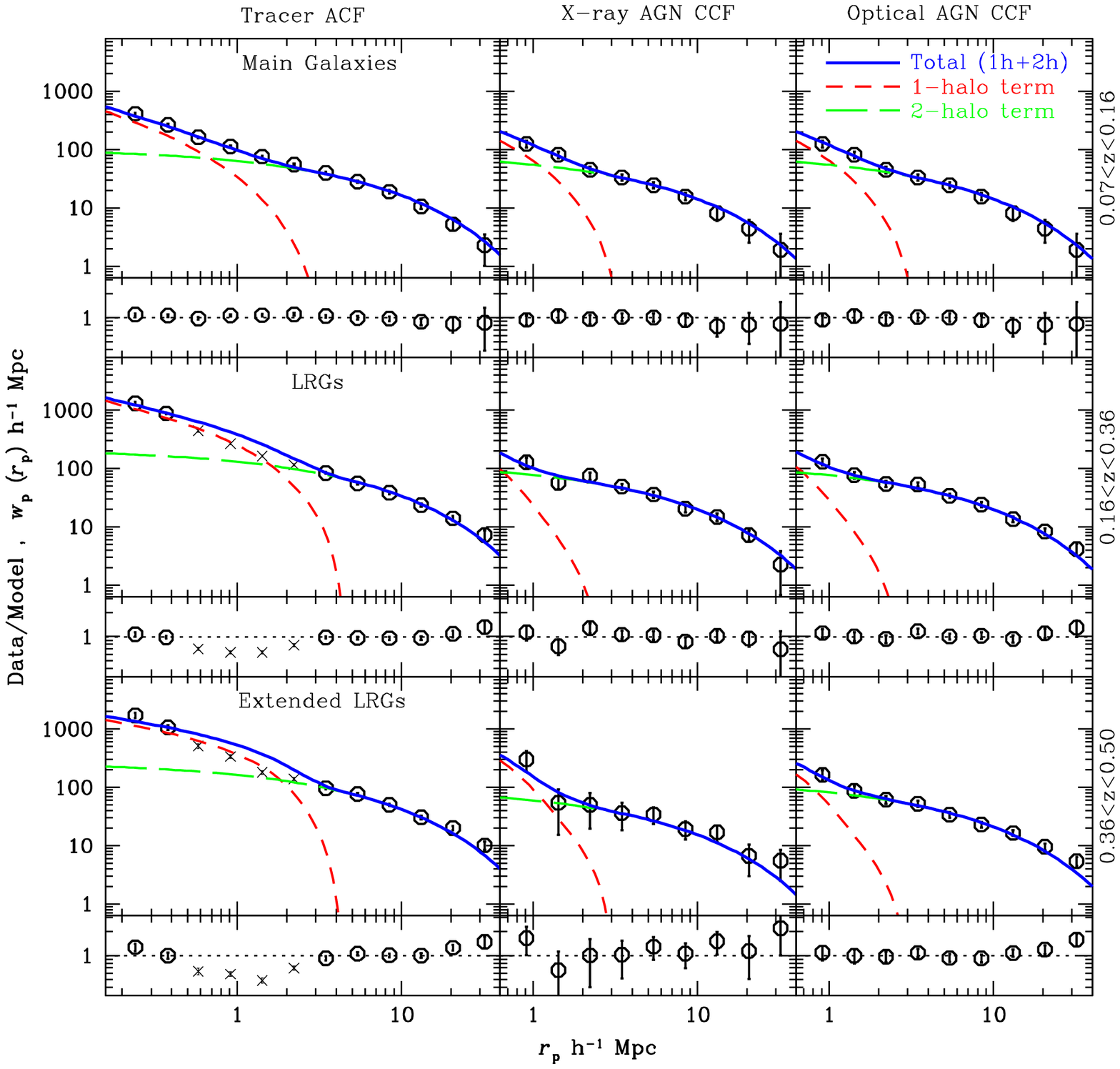}} 
      \caption{Projected tracer set ACFs and CCFs between the different total 
               AGN samples and tracer sets in the corresponding redshift ranges.
               The order of the panels is the same as in Fig.~\ref{contour_acf}.
               Each ACF/CCF plot contains the corresponding best-fit HOD model 
               (blue solid line) along with the one halo (red dashed line) and two halo 
               (green long-dashed line) terms and shows the residuals
               (data/model) below each individual plot. We show the $r_{\rm p}$ range 
               that is used for the fits, i.e., $r_{\rm p}>0.2$ $h^{-1}$ Mpc and 
               $r_{\rm p}>0.7$ $h^{-1}$ Mpc for the tracer ACFs and the AGN-tracer CCFs
               respectively. The LRG ACF data points shown as crosses are not used in 
               the fit (see Sect.~\ref{sec:lrghod}).}     
         \label{wp_CCF_AGN}
\end{figure*}

\subsubsection{SDSS Main Galaxies ($0.07<z<0.16$)}
As explained above in Section~\ref{MainGal}, the tracer set for the low redshift AGN samples is
selected from the SDSS main galaxy sample with an absolute magnitude range of 
$-22.1 < M^{0.1}_r < -21.1$. The number density of this sample is 
$(9.77\pm 0.16)\times 10^{-4}$ $h^{3}$ Mpc$^{-3}$. We use the five-parameter model 
by \citet{zheng_coil_2007} to represent the central and satellite HODs of our low
redshift tracer set:
\begin{eqnarray}
\langle N_{\rm G,c}\rangle (M_{\rm h}) &=& 
\frac{1}{2}\left[1+\mathrm{erf}\left(\frac{\log M_{\rm h}-\log M_{\rm
min}}{\sigma_{\log M}}\right)\right]\nonumber\\
\langle N_{\rm G,s}\rangle (M_{\rm h}) &=& \langle N_{\rm G,c}\rangle (M_{\rm h})\;
\left(\frac{M_{\rm h}-M_0}{M_1^\prime}\right)^{\alpha_{\rm s}}.
\label{eq:zheng07hod}
\end{eqnarray}

This form involves a step function with a lower mass cutoff $M_{\rm min}$, 
which is smoothed by incorporating the error function (erf) with the width of the cutoff 
profile $\sigma_{\log M}$. For the detailed description of 
the different parameters, see Sect.~3.2 of \citet{zheng_coil_2007}. 
A limitation of our current fitting software is that it allows a maximum of two
simultaneous variable parameters. Thus, we search for acceptable fits in a two
parameter space, while fixing other parameters to reasonable values. We note that, for our current
purposes in this paper of obtaining correct $b_{\rm A}$ values, which is mainly constrained by the
two halo term, the detail of the HOD is not critical (see discussions below in
Sect.~\ref{sec:agn_bias_mdmh}). A Code that searches for the best-fit
values and confidence ranges of model parameters in a many-parameter space using the 
Markov-Chain Monte-Carlo (MCMC) is under development and will be used in a future paper.

\citet{zheng_coil_2007} perform fits to the full five parameter space for several 
luminosity-threshold subsets of the SDSS main galaxy sample. We take advantage of 
the results of their $M_{\rm r}<21.0$ sample, which has a similar selection criteria to our SDSS main galaxy 
sample, and use their results to fix three parameters: $\log (M_0/[h^{-1}\,M_{\sun}])=11.92$,
$\log (M_1^\prime/[h^{-1}\, M_{\sun}])=13.94$, and $\sigma_{\log M}=0.39$. We then search for 
the best fit model for our measurement of the SDSS main galaxy ACF in the 
remaining two parameters ($\log M_{\rm min}$ and $\alpha_{\rm s}$) by minimizing the correlated 
$\chi^2$, taking into account the density constraint (see Eq.~18 of paper II). We find an excellent 
fit to the data with the best fit values and 1$\sigma$ uncertainties of 
$\log (M_{\rm min}/[h^{-1}\, M_{\sun}])=12.81\pm 0.01$ and $\alpha_{\rm s}=1.16\pm 0.02$, with an 
associated bias parameter of $b_{\rm G}=1.43\pm 0.01$. 
The uncertainties in these values are lower than those 
in \citet{zheng_coil_2007} 
as we perform the fit with only two free parameters instead of five.
For the fit, we use $w_{\rm p}(r_{\rm p})$ 
measurements in the range $0.2 <r_{\rm p}[h^{-1}\,{\rm Mpc}]<40$.  First, we follow our 
HOD modeling described in paper II and exclude data points that fall in the 
transition region between the one halo dominated and two halo dominated regimes.
In the case of the SDSS main galaxy sample, using scales of 0.4 $<r_{\rm p}[h^{-1}\,{\rm Mpc}]<$ 1
 results in essentially the identical best-fit HOD. The SDSS main galaxy ACF and the 
best fit model are shown in Fig.~\ref{wp_CCF_AGN} (upper left panel).

\subsubsection{Luminous Red Galaxies ($0.16<z<0.36$)}  
\label{sec:lrghod}
The tracer set in the intermediate redshift range is the LRG sample with 
$-23.2<M_g^{0.3}<-21.2$ (Sect.~\ref{desc_LRG}). The derivation of central and 
satellite HODs of this sample is described in detail in paper II. In summary,
we start from the results of \citet{zheng_zehavi_2009} and make a two parameter adjustment 
to find the HOD that fits best to our LRG ACF, including the number density constraint. 
For satellite LRGs, we interpolate between the $\Delta \chi^2=4$ upper and lower bounds 
of the satellite HOD by \citet{zheng_zehavi_2009}, while for the central LRG HOD we shift 
their central HOD horizontally along the $\log M_{\rm h}$ axis. First, we fit 
$w_{\rm p}(r_{\rm p})$ measurements in the $0.2<r_{\rm p}\,[h^{-1}\,{\rm Mpc}]<40$ range.
Significant residuals remain in the transition region between the one halo and
two halo term dominated regimes, due to the fact that our HOD modeling neglect the effects of
halo-halo collisions in the two halo term. Unlike with the main galaxy sample, this effect is not 
negligible with the LRG sample, due to a larger transition region. We therefore 
neglect data points in a range of 
$0.46<r_{\rm p}\,[h^{-1}\,{\rm Mpc}]<2.8$ and find a good fit to the data 
(Fig.~\ref{wp_CCF_AGN}, middle left panel). The associated LRG bias parameter 
is $b_{\rm G}=2.20\pm 0.01$, where this error includes only the statistical 1$\sigma$ 
uncertainty of the fit.

\subsubsection{Extended LRGs ($0.36<z<0.50$)}
Unlike for the cases of the SDSS main galaxy and LRG samples, there is no
template HOD model in the literature for a sample with almost identical 
selection criteria to those for our extended LRG sample, i.e., a non-volume 
limited sample of LRGs with $-23.2<M_g^{0.3}<-21.7$ at $0.36<z<0.50$.

The closest sample that we can use as our template is the 
$-23.2<M_g^{0.3}<-21.8$, $z\sim 0.3$ LRG sample, for which
\citet{zheng_zehavi_2009} made a detailed HOD investigation, in addition to the 
$-23.2<M_g^{0.3}<-21.2$ LRG sample. Thus, we follow the approach in paper II 
for the 
LRG sample and search for the best-fit HOD model of the ACF of our 
extended LRG sample by adjusting \citet{zheng_zehavi_2009} HOD results for the 
$-23.2<M_g^{0.3}<-21.8$ LRG sample.

In short, we take the central and satellite HODs from Fig.~1 (b) of \citet{zheng_zehavi_2009}
and search for the best-fit HOD by tweaking the template. We shift their central HOD 
horizontally by $d$ in $\log M_{\rm h}$ and interpolate between their upper and lower 
bounds ($\Delta \chi^2$) of their satellite HOD, with the dividing ratio of $f:(1-f)$,
where $f=0$ ($f=1$) represents their lower (upper) bound on the satellite HOD (see paper II 
for details). We search for the best-fit HOD model in the $(f,d)$ space, 
where the HOD-model predicted $w_{\rm p}(r_{\rm p})$ function is calculated at $z=0.42$.
As with the LRGs, we exclude the $0.46<r_{\rm p} [h^{-1}\,{\rm Mpc}]<2.8$ 
range from the fit. We obtain the best-fit with $f=4.2\pm 0.8$ and $d=-0.29\pm 0.04$. 
We do not use the density constraint in the fit, because the extended LRG sample is not 
volume-limited and therefore the number density is not accurately determined. However, 
the best-fit model gives a density of $4.3\times 10^{-5}$ $h^{3}$ Mpc$^{-3}$, which is close 
to the number density of the extended LRGs calculated in the volume-limited portion of the 
sample (see Table~\ref{samples}).

\subsection{AGN biases and Typical Halo Masses}
\label{sec:agn_bias_mdmh}

For the AGN HOD model, we use model B of paper II, which parameterizes the number 
of central ($\langle N_{\rm A,c}\rangle$) and satellite ($\langle N_{\rm A,s}\rangle$) 
AGNs in a DMH:
\begin{eqnarray}
\langle N_{\rm A,c}\rangle &=& f_{\rm A}\Theta(M_{\rm h}-M_{\rm min}), \nonumber \\
\langle N_{\rm A,s}\rangle &=& f_{\rm A}\Theta(M_{\rm h}-M_{\rm min})(M_{\rm
h}/M_1)^{\alpha_{\rm s}},  
\label{eq:step_pl}
\end{eqnarray}
where  $\Theta(x)$ is the step function (equal to 1 at $x\geq 0$; 0 at $x<0$) and
$f_{\rm A}$ represents the AGN fraction (duty cycle) among central galaxies at 
$M_{\rm h}\ga M_{\rm min}$. We use $\log M_1/M_{\rm min}=1.36$ in
Eq.~\ref{eq:step_pl}, which \citet{zehavi_zheng_2005} find to be a typical value 
at which a DMH hosts on average one satellite galaxy in addition to a central galaxy.  
We model the CCFs between our tracer sets and the 
AGNs using the same method as in paper II:
\begin{eqnarray}
P_{\rm AG,1h}(k)&=&\frac{1}{(2\pi)^3n_{\rm A}n_{\rm G}}\int\phi(M_{\rm
h})\times\nonumber\\
        & & [\langle N_{\rm A,c}N_{\rm G,s}+N_{\rm A,s}N_{\rm G,c}\rangle(M_{\rm
h})\,\,y(k,M_{\rm h})+\nonumber\\
        & & \langle N_{\rm A,s}N_{\rm G,s}\rangle(M_{\rm h})\,\,|y(k,M_{\rm h})|^2]\,
        dM_{\rm h}, \label{eq:pag1h}\\ \nonumber\\
P_{\rm AG, 2h}(k) &\approx& b_{\rm A}b_{\rm G}  P_{\rm lin}(k), 
\label{eq:p_ccf_2h}
\end{eqnarray}
and
\begin{equation}
w_{\rm p}(r_{\rm p})=\int \frac{k}{2\pi}[P_{\rm 1h}(k)+P_{\rm 2h}(k)]J_0(kr_{\rm
p})dk 
\label{eq:wp_pk}
\end{equation}
where $n_{\rm A}$($n_{\rm G}$) is the number density of AGNs (tracers), 
$y(k,M_{\rm h})$ is the Fourier transform of the Navarro, Frenk, \& White 
profile (NFW; \citealt{navarro_frenk_1997}), $P_{\rm lin}(k)$ is the linear power 
spectrum of the density field with the transfer function by \cite{eisenstein_hu_1998}, 
and $J_0(x)$ is the zeroth-order Bessel function of the first kind. 

Due to the low space density of AGNs, most CCFs do not have a signal-to-noise ratio
on small scales that is sufficient for applying the $\chi^2$ statistics. 
While a majority of CCFs contain $\ga 16$ pairs per bin at $r_{\rm p}>0.3$ $h^{-1}$ Mpc, 
we have to exclude $r_{\rm p}<0.7$ $h^{-1}$ Mpc bins for several CCFs in order to
have at least 16 AGN-galaxy pairs per bin. To derive $b_{\rm A}$ in a consistent way for
all CCFs, we use only $r_{\rm p}>0.7$ $h^{-1}$ Mpc bins to derive column `$b(z)$ HOD' for
all of the AGN samples given in Table~\ref{xagn_acf}.

Since the purpose of using the HOD modeling in this paper is to derive reliable 
values of $b_{\rm A}$, we do not discuss detailed results using other models, which
will presented in Miyaji et al. (in preparation).  The derived bias parameter, which is 
mainly constrained by the two halo term, is not very sensitive to our particular 
choice of HOD model.
To verfiy this, we repeat the HOD fits to the CCFs with different values
of the parameter $\log M_1/M_{\rm min}$ (Eq.~\ref{eq:step_pl}). For various SDSS 
luminosity-threshold galaxy samples, \citet{zehavi_zheng_2005} found the range of this parameter 
to be $1.0\la \log M_1/M_{\rm min}\la 1.5$. Thus we fix $\log M_1/M_{\rm min}$
to 1.0, 1.36 (our default), and 1.6 and obtain $b_{\rm A}$ values in each case. 
The best fit values of $b_{\rm A}$ typically vary only by $\approx 0.01$ among these three cases, 
demonstrating the robustness of deriving the bias parameter using this method.

\subsection{AGN HOD bias results} 

\begin{figure}
  \centering
 \resizebox{\hsize}{!}{ 
  \includegraphics[bbllx=56,bblly=362,bburx=530,bbury=706]{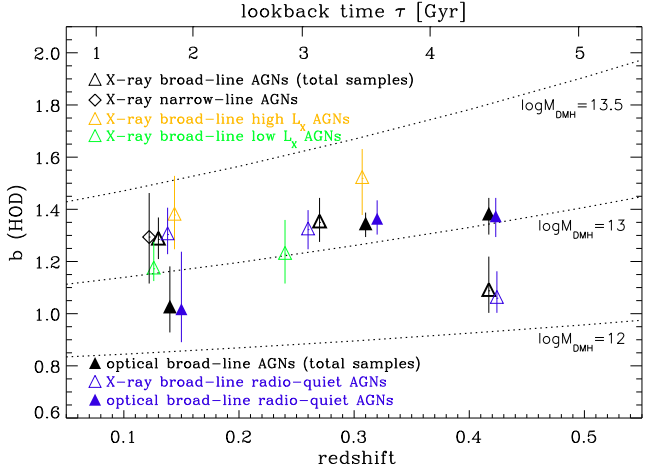}}
      \caption{HOD modeling bias parameter $b_{\rm AGN}$ as a function of redshift for our 
                 X-ray selected RASS/SDSS samples (open symbols) and optically-selected SDSS 
                 AGN samples (filled symbols). Black symbols represent the total AGN 
                 samples, while colored symbols represent subsamples (low and high 
                 $L_{\rm X}$ in green and yellow, radio-quiet in blue). The dotted lines 
               show the expected $b(z)$ of typical DMH masses $M_{\rm DMH}$ based on \cite{sheth_mo_2001}
               and \cite{tinker_weinberg_2005}, where masses are given in log $M_{\rm DMH}$ in units of 
               $h^{-1}$ $M_\odot$. For visualization purposes, we slightly shift the redshifts of some AGN samples.}
              \label{xray_optical_b_HOD}
\end{figure}

In Fig.~\ref{xray_optical_b_HOD} we present the main results of our study. 
We show the HOD bias parameter for our different X-ray and optically-selected AGN 
samples as a function of redshift. All AGN samples are consistent with a host 
DMH mass of log $(M_{\rm DMH}/[h^{-1}\,M_{\odot}]) \sim 12.4-13.4$ 
(see Table~\ref{xagn_acf}). Samples in which the radio-loud AGNs have been
excluded have very similar clustering amplitudes as the total samples, in 
all three redshift ranges. The clustering signal of narrow-line 
RASS/SDSS AGNs at $0.07<z<0.16$ is also very similar to broad-line RASS/SDSS 
AGNs at the same redshift. 
Furthermore, weak X-ray luminosity dependences on the broad-line AGN 
clustering amplitude are found at both $0.07<z<0.16$ and $0.16<z<0.36$.

\subsection{Power Law versus HOD Derived Bias Parameters}
\begin{figure}
  \centering
 \resizebox{\hsize}{!}{ 
  \includegraphics[bbllx=84,bblly=370,bburx=544,bbury=696]{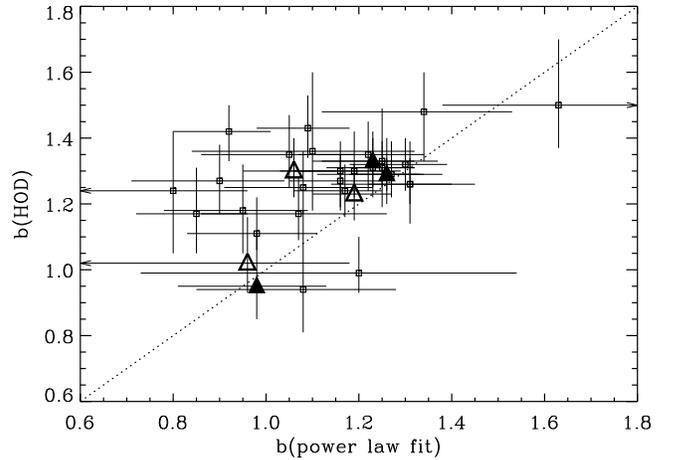}} 
      \caption{Comparison between bias parameters derived from power law fit 
               versus HOD modeling. 
                We highlight the most important samples at the three 
                redshift ranges: total X-ray selected RASS/SDSS AGN samples 
                (open triangles) and total optically selected SDSS AGN samples 
                (filled triangles).
                The 1$\sigma$ uncertainties for the 
                different methods are plotted as error bars. The dotted line 
                shows a 1:1 correspondence.}
         \label{compare_b}
\end{figure}

The various AGN samples studied here allow us to compare 
the power law fit derived bias parameters with those from HOD modeling. 
In Fig.~\ref{compare_b} we plot the bias values from the power law fits and HOD modeling 
listed in Table~\ref{xagn_acf}. The $b$(HOD) values are often larger than the $b$(power law fit) 
values. Furthermore, the uncertainties on the HOD bias are often lower than on the power law bias.
This is because the power law fits are based on the inferred ACF, while 
the HOD modeling is directly applied to the CCF which has lower statistical uncertainties.

The relatively narrower distribution and on-average 
higher $b$(HOD) values compared to the $b$(power law fit) values 
are caused by strong variations between the samples in the one halo term, 
while the variations in the two halo term are smaller.
As described in the beginning of Section~\ref{hod_model}, power law fit bias 
measurements commonly use smaller scales that are in the one halo term 
(our fitted range is $r_p=0.3-15$ $h^{-1}$ Mpc) in order to increase the 
statistical significance. If power law fits are restricted only to larger scales, 
the method suffers
from the problem that the lowest scale where the linear biasing scheme can still be 
applied varies from sample to sample and remains ambiguous. HOD modeling allows, 
in principle, the use of the full range of scales since the method first determines the 
one and two halo terms and then constrains the linear using data 
down to the smallest $r_p$ values that are dominated by the two halo term for each 
individual sample.

Therefore, an apparently low power law bias value of one sample compared to 
another can be due to a lower clustering strength in the one halo term, while 
the ``true'' linear bias parameters may be similar. If we use scales of 
$r_{\rm p}>1.5$ $h^{-1}{\rm Mpc}$ for the power law fits, the errors on the bias 
are much larger, but the values are
statistically consistent which those derived from the HOD model fits.   

In summary, on small ($\lesssim 1 h^{-1}$ Mpc) and large ($\gtrsim 10 h^{-1}$ Mpc) 
scales a power law model is not a good fit to the data. 
HOD modeling is currently the optimal method to establish the large-scale
bias parameter, provided the adopted HOD parameterization can adequately describe 
the true galaxy and AGN DMH population.
With larger data 
sets in the future, the uncertainties in the measurements will be decreased further. 
To derive a reliable picture of AGN clustering, bias parameters should be inferred from 
HOD modeling, or at least from the comparison of the correlation function with that of 
the dark matter only in the linear regime \citep{allevato_2011}, because systematic errors 
based on power law bias parameters will be larger than the statistical uncertainties 
of the clustering measurement. In the following discussions, we use the HOD model bias 
parameters whenever we compare our different AGN samples. In the case where we compare 
our results to other studies, which list only power law fit bias values, we also use power 
law fit bias values to be consistent.


\section{Discussion}

Our clustering measurements of luminous broad-line AGNs yield three independent 
data points in the poorly-studied low-redshift range for both X-ray and optically-selected broad-line AGNs.
In addition, we measure the clustering signal of X-ray detected narrow-line AGNs. We derive the 
bias parameter of the different samples based on power law fits and HOD modeling.

\subsection{Comparison with X-ray selected broad-line AGN Clustering Measurements}
There are no published clustering measurements of X-ray selected broad-line AGNs 
with comparable low uncertainties at low redshifts. For example, \cite{mullis_henry_2004}
measure the clustering strength of broad-line AGNs in the {\em ROSAT} NEP survey, for which we derive a 
bias parameter of $b=1.83^{+1.88}_{-0.61}$ ($z=0.22$, paper I).

At higher redshift \cite{allevato_2011} compute the clustering strength of X-ray (0.5--2 keV) 
unabsorbed \& absorbed and narrow \& broad-line AGN in the {\em XMM-Newton} COSMOS field 
in different bins where the median redshift of the subsamples varies from $z=0.5$ to
$z=2.5$. They find that broad-line AGNs reside in DMH of 
log $(M_{\rm DMH}/[h^{-1}\,M_{\odot}]) \sim 13.2$, independent of redshift. The average luminosity
of their broad-line AGN sample is log ($L_{\rm 2-10\,{\rm keV}}/{\rm [erg s^{-1}]}) \sim 44.1$ 
(intrinsic absorption-corrected luminosity; V. Allevato, private communication 2011)
which corresponds to  log ($L_{\rm 0.1-2.4\,{\rm keV}}/{\rm [erg s^{-1}]}) \sim 44.2$ assuming $\Gamma=1.8$. Consequently, 
the broad-line AGNs studied by \cite{allevato_2011} have X-ray luminosities comparable to our total RASS/SDSS 
AGN sample at $0.16<z<0.36$. The derived DMH mass of log $(M_{\rm DMH}/[h^{-1}\,M_{\odot}]) \sim 13.2$ 
for our low-redshift RASS/SDSS AGN sample is in excellent agreement with the value found at $z\sim 1.5$ for the broad-line 
AGN sample of \cite{allevato_2011}.

\subsection{Optical versus X-ray selected AGN clustering properties}
\label{discussion_optical_xray}
\begin{figure*}
  \centering
 \resizebox{\hsize}{!}{ 
  \includegraphics[bbllx=72,bblly=360,bburx=530,bbury=706]{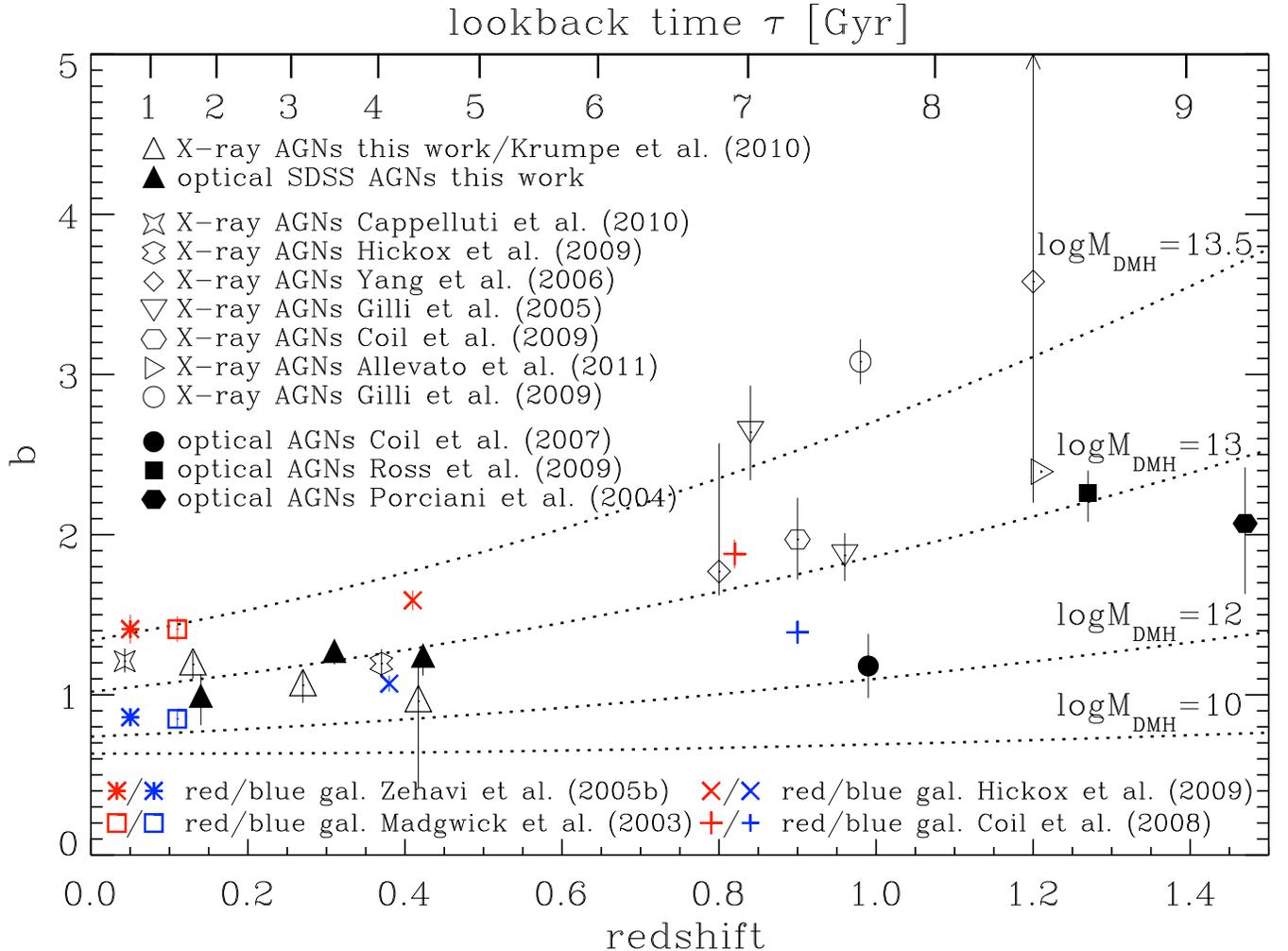}}
      \caption{Power law bias parameter $b_{\rm AGN}=\sigma_{8,\rm AGN}(z)/\sigma_8D(z)$ as a 
               function of redshift for various X-ray and optically-selected AGN samples as well 
               as blue and red galaxies. We plot our power law fit bias parameters to compare different 
               studies in a consistent manner. Black open symbols represent X-ray selected AGN samples, 
               while black filled symbols represent optically-selected AGN samples. 
               Clustering measurements for red and blue galaxies 
               are shown as red and blue symbols at different redshifts. For the explanation of the 
               dotted lines see Fig.~\ref{xray_optical_b_HOD}.
               For visualization purposes, 
               we slightly offset the redshifts of the X-ray and optically-selected SDSS AGN samples.}
              \label{xray_optical_b}
\end{figure*}

Our study allows us to compare X-ray and optically-selected AGN clustering measurements over three 
independent low redshift ranges. As the same procedure and identical tracer sets are used 
for calculating the clustering strengths of the X-ray and optically-selected AGN samples, we expect 
there to be minimal systematic errors when comparing the clustering properties among 
our different AGN samples. 

Since most of the results in the literature are based on power law fits to the correlation 
function, we compare our power law bias parameters from various clustering measurements of X-ray and 
optically-selected SDSS AGN samples and galaxy clustering measurements in 
Fig.~\ref{xray_optical_b}.
The bias values of our X-ray and optically-selected 
AGN populations clearly fall within the region occupied by $\sim$$L^*$ galaxies. 

We do not find a very significant difference between the clustering amplitude of 
our broad-line optical and X-ray AGN samples in the different redshift ranges considered here.
The differences in the HOD model bias parameters between the total X-ray and optically-selected  
AGN samples in our three (increasing) redshift ranges are 1.5$\sigma$, 0.1$\sigma$, and 2.0$\sigma$. 
As our samples at $0.16<z<0.36$ have many more X-ray and optically-selected AGNs than the 
samples at lower and higher redshifts, we consider the results in this redshift range to
be the most reliable.  We do not find a significant difference in the clustering of X-ray 
and optically-selected AGN in this redshift range. Moreover, given that only one out of three 
sample has a difference of $\sim$2.0$\sigma$, we conclude that it is likely not significant. 

Although the RASS/SDSS AGN samples extend to 
lower AGN luminosities than the optically-selected SDSS AGNs, both samples span roughly the same 
luminosity range for broad-line AGNs, which may account for the similar clustering properties. 
RASS and SDSS are also well matched in terms of the depth and selection of broad-line AGNs 
(see \citealt{anderson_voges_2003, anderson_margon_2007}). Optical surveys are known to find 
at brighter magnitudes predominantly X-ray unabsorbed broad-emission line AGNs.
{\em ROSAT}'s soft energy 
range allows primarily for the detection of X-ray unabsorbed AGN and, hence, is also biased  
also toward broad-line AGNs. Therefore, we do not expect strong systematic differences between
these sampes due to the RASS/SDSS AGN selection. 

For the optically-selected SDSS broad-line AGNs in the redshift ranges of $0.16<z<0.36$ and $0.36<z<0.50$, 
the sample is large enough to create subsamples of optically-selected AGN that are not detected in RASS. 
Furthermore, at $0.16<z<0.36$ we select optically-selected AGNs that are 
also detected by RASS. We use these samples as a consistency check to verify that the clustering 
properties between X-ray and optically-selected AGN samples are not significantly different. 
As shown in Table~3 all of these AGN samples agree well within their 1$\sigma$ uncertainties in their 
HOD model bias parameters. This provides compelling evidence that there is indeed no strong difference 
between broad-line X-ray and optically-selected AGNs at similar luminosities at low redshifts.

Figure~\ref{xray_optical_b} compares our AGN clustering results to other X-ray and 
optically-selected AGN clustering studies. The properties of various clustering studies are given 
in Table~4 of paper I. \cite{hickox_jones_2009} study the clustering properties of AGNs in the 
AGES survey. As they publish only the AGN CCFs with galaxies, 
we use their best power law fits (R. Hickox, private communication 2011) to infer their AGN ACF 
and the bias parameter, following our approach described in Sect.~\ref{inferringACF}. 
 From the redshift distributions presented in 
\cite{hickox_jones_2009}, we compute the effective median redshift for the CCFs ($\overline{z_{\rm eff}}=0.37$) 
and derive a bias value of $b_{\rm X,Hi09}=1.20^{+0.09}_{-0.09}$.

Our finding of detecting no significant difference in the AGN clustering properties between X-ray 
and optically-selected AGNs at low redshifts may appear to be in contrast to AGN clustering measurements 
at higher redshifts ($z>0.7$), where optically-selected AGN samples have a lower clustering strength than 
X-ray selected AGN samples (Fig.~\ref{xray_optical_b}). Furthermore, the clustering of X-ray selected AGNs 
is roughly consistent with red galaxies at higher redshifts. However, some of the X-ray clustering studies
significantly underestimate their uncertainties by using Poisson errors instead of jackknife errors.
Moreover, X-ray and optically-selected AGN samples at these redshifts select AGNs with different 
intrinsic properties. While the optical AGNs are mainly drawn from large sky area surveys and 
sample luminous predominantly broad-line AGNs, the X-ray selected AGN samples derive 
from very deep observations covering only a few square degrees on the sky. Consequently, the X-ray 
selected AGNs have, on average, much lower luminosities. Additionally, the X-ray samples include
absorbed AGNs, which results in a large fraction of narrow-line AGNs that are missed in the optical 
AGN samples at these redshifts.

\subsection{Broad-line versus Narrow-line AGNs}
The differences in the clustering signals between X-ray and 
optically-selected AGN samples at $z>0.7$ can potentially be accounted for either as the result of 
a large luminosity difference between the samples or the large fraction of X-ray absorbed, optically 
narrow-line AGNs in the X-ray AGN samples.

To test the latter assumption, we measure the clustering properties of narrow-line RASS/SDSS 
AGNs classified by \cite{anderson_margon_2007} in the redshift ranges of 
$0.07<z<0.16$ and $0.16<z<0.36$. The low number density of narrow-line RASS/SDSS AGNs at $0.16<z<0.36$
forces us to use scales of $r_P>1.1$ $h^{-1}$ Mpc to apply the $\chi^2$-statistics during 
the HOD modeling. The clustering strength of the narrow-line and total broad-line RASS/SDSS AGN samples
at $0.07<z<0.16$ agree well with each other (Table~\ref{xagn_acf}), although the uncertainty 
for the narrow-line AGN sample 
is large. The power law fit bias parameter for the narrow-line RASS/SDSS AGN sample 
is significantly lower than the bias derived from the HOD model (see Fig.~\ref{xray_optical_b_HOD}) 
and would suggest a significantly  
lower clustering amplitude than the broad-line RASS/SDSS AGN sample. This is caused by differences 
between the samples on small scales where the one halo term dominates. The power law fits use 
these small scales, while the HOD modeling mainly relies on the two halo term to determine the large-scale 
clustering.

At $0.16<z<0.36$ the narrow-line AGN HOD 
bias parameter is too poorly constrained ($b=1.01^{+0.24}_{-0.17}$) to allow for a detailed interpretation.    
While we cannot determine whether narrow-line RASS/SDSS AGNs cluster similarly or less than broad-line 
AGNs in that redshift range, we can rule out the conclusion that they are significantly more clustered. 
Two important points are worth noting: first, as {\it ROSAT} can detect 
only moderately X-ray absorbed AGNs (log $(N_{\rm H}/[{\rm cm}^{-2}])\la 22$), while {\it XMM-Newton} and 
{\it Chandra} are sensitive to much more absorbed (and lower luminosity) AGNs, the narrow-line AGNs 
detected by RASS may not be as common. Second, the relative clustering strength of narrow-line AGNs 
could change with redshift if there is a difference in how AGN activity is triggered at different 
cosmological epochs.  However, other studies confirm that low-redshift narrow-line (radio-quiet) AGNs 
are not strongly clustered and are hosted in galaxies that do not differ significantly from 
typical non-AGN galaxies.(e.g, \citealt{mandelbaum_li_2009}, \citealt{li_kauffmann_2006}).

\cite{allevato_2011} find that X-ray selected narrow-line AGNs in the {\em XMM-Newton} COSMOS field
cluster slightly lower (2.3$\sigma$) than X-ray selected broad-line AGNs and reside in DMHs of 
log $(M_{\rm DMH}/[h^{-1}\,M_{\odot}]) \sim 13.0$ in the redshift range $z\sim 0.5-1.0$. 
However, their narrow-line AGNs have an average intrinsic (absorption corrected) 
log ($L_{\rm 2-10\,{\rm keV}}/{\rm [erg s^{-1}]}) = 43.1$ (V. Allevato, private communication 2011) 
which is an order of magnitude lower than the average luminosity of their broad-line AGNs.
When \cite{allevato_2011} consider only the X-ray properties of the sources to create  
X-ray absorbed and X-ray unabsorbed subsamples, in which both have an almost identical mean 
luminosity of log ($L_{\rm 2-10\,{\rm keV}}/{\rm [erg s^{-1}]}) \sim 43.65$, they find that  
X-ray absorbed AGNs cluster slightly less (2.6$\sigma$) than X-ray unabsorbed AGNs. 

To summarize, the difference between the AGN clustering properties between X-ray and 
optically-selected AGN samples at $z>0.7$ is likely not due to a strongly clustered  
population of narrow-line AGNs in the X-ray samples. However, these objects do have 
significantly lower luminosities than optically-selected broad-line AGNs.

\subsection{Impact of Radio-loud Broad-line AGN on the Clustering Signal}
For each of the various broad-line AGN samples studied here, we create subsamples where we have excluded 
radio-detected AGNs to study the impact of radio-loud AGNs on the derived clustering strength.
We do not find any significant differences between the AGN samples that include or 
exclude radio-detected AGNs. The HOD bias parameters for these samples
 agree well within their 1$\sigma$ uncertainties in all three redshift 
ranges studied (see Fig.~\ref{xray_optical_b_HOD}). However,  the samples 
are similar as only approximately 10--20\% of all broad-line AGNs have also radio (FIRST) detections,
so this result may not be particularly constraining.

As mentioned in Section~\ref{optically_selected_desc}, the SDSS AGN target selection 
gives the highest priority to point sources that are detected in FIRST (above a certain 
flux limit) without considering their colors. We select all AGNs ($n=187$) with $0.16<z<0.36$ which have 
the SDSS FIRST target selection flag equal to 1. Hence, these objects were only selected on the basis 
of having a significant FIRST radio flux and can be understood as a well-defined radio-selected AGN sample 
($\langle$$z$$\rangle$=0.28, $\langle$$M_i$$\rangle$=-23.28). We compute the CCF of this AGN sample 
with the LRG tracer set. 
Due to the low number of AGNs in this sample, the HOD 
bias parameter has large uncertainties, with a value of $b=1.45^{+0.24}_{-0.29}$. A sample of all 423 
SDSS AGNs that have a radio-detection (including the 187 radio-selected SDSS AGNs; $\langle$$z$$\rangle$=0.27, 
$\langle$$M_i$$\rangle$=-22.84) yields $b=0.97^{+0.18}_{-0.19}$. 
Furthermore, we compute the CCF of only the RASS/SDSS AGNs with $0.16<z<0.36$  
that are marked as radio sources in \cite{anderson_margon_2007}. The sample, which contains 144 objects and 
has $\langle$$z$$\rangle$=0.25 and log $\langle$$L_{\rm 0.1-2.4\,{\rm keV}}/{\rm [erg s^{-1}]})$$\rangle$$ \sim 44.36$,
yields a HOD bias parameter of  $b=1.08^{+0.30}_{-0.32}$. 
No constraining results can be drawn by using only radio-detected SDSS AGN samples, because these samples 
contain too few objects. However, all values are consistent 
with the clustering strengths of the other AGN samples in the same redshift range.

\begin{figure}
  \centering
 \resizebox{\hsize}{!}{ 
  \includegraphics[bbllx=60,bblly=362,bburx=537,bbury=705]{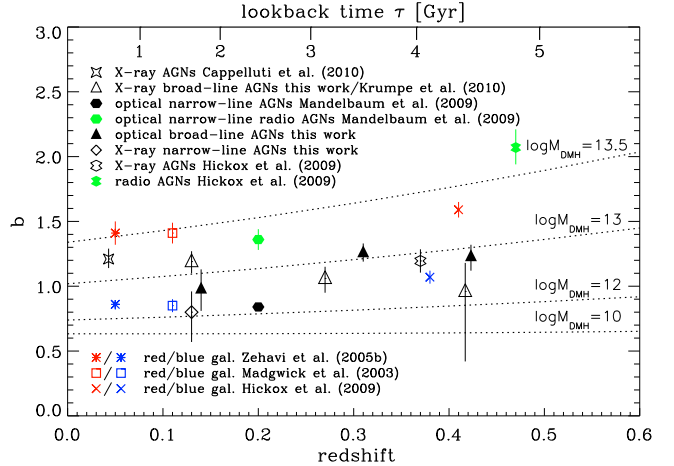}}
      \caption{Power law bias versus redshift (similar to Fig.~\ref{xray_optical_b_HOD}), comparing 
               the clustering of differently-selected AGN samples (radio and non-radio samples) and 
               galaxies. Note that the 
               HOD model bias parameter for the narrow-line RASS/SDSS AGN 
               sample is significant higher ($b=1.24^{+0.18}_{-0.19}$) than the power law bias shown here.}
              \label{lowz_b}
\end{figure}

Many previous studies (e.g., \citealt{magliocchetti_maddox_2004}; \citealt{hickox_jones_2009}; 
\citealt{mandelbaum_li_2009}) found that radio-loud AGNs cluster more strongly than AGNs 
without the presence of radio emission.
At first glance our results may appear to be in contrast to these findings. 
However, those studies focus on the clustering properties of optical 
narrow-line (instead of broad-line) AGNs based on diagnostic emission-line ratios and 
radio luminosity. Moreover, when comparing the clustering strength of different samples 
it is essential to take into account the involved luminosities.
Figure~\ref{lowz_b} shows the bias of various AGN and galaxy samples
at lower redshift ($z \le 0.6$), focusing on comparing radio-selected AGN samples.
For the radio (non-broad-line) AGN sample of \cite{hickox_jones_2009}, we derive a bias value
of $b_{\rm radio,Hi09}=2.07^{+0.14}_{-0.13}$ ($\overline{z_{\rm eff}}=0.47$) by following 
the description given in Sect.~\ref{discussion_optical_xray}. 

\cite{magliocchetti_maddox_2004} find that radio (FIRST) galaxies with AGN activity cluster more 
strongly than both radio-detected 2dFGRS galaxies without AGN activity and 2dFGRS galaxies without radio 
emission. It is not clear, however, whether the different galaxy samples have similar 
luminosities and colors.
If the galaxy samples that are used for comparison are mainly blue, star-forming galaxies, one would
expect a significant difference in the clustering strength based on the different host galaxy 
properties. Consequently, the clustering difference could reflect not by AGN activity but 
host galaxy type. The high bias parameter for radio-loud AGNs ($b=2.07^{+0.14}_{-0.13}$) 
and moderate bias ($b=1.20^{+0.09}_{-0.09}$) for 
X-ray selected AGNs in \cite{hickox_jones_2009} can also be explained by the different 
host galaxy populations (see their Fig.~9).

\cite{mandelbaum_li_2009} compare the clustering strength of optically-selected narrow-line 
AGNs ($0.01<z<0.3$ and $-23< M_{r}^{0.1}<-17$) with and without radio emission at {\it fixed}
stellar mass, i.e., at fixed luminosity and color. They find that at fixed stellar mass 
radio-loud narrow-line AGNs cluster more strongly and hence reside in more massive DMHs 
than narrow-line AGNs without radio emission and galaxies without the presence of AGN 
activity (see Fig.~\ref{lowz_b}). Neither \cite{magliocchetti_maddox_2004} nor 
\cite{mandelbaum_li_2009} detect a difference in the clustering properties within the 
radio-loud narrow-line AGN sample as a function of radio luminosity.

\cite{shen_strauss_2009} compute the ACF of optical SDSS DR5 AGNs 
(from the \citealt{schneider_hall_2007} sample) in the redshift range
 $0.4 \lesssim z \lesssim 2.5$.
As we do here, they divide their samples into FIRST-detected and undetected sources. 
Their fits (over the range 3--115 $h^{-1}$ Mpc) suffer large uncertainties; therefore, 
they fix the slope of the power law fit for all samples to $\gamma=2$. They compute 
the ACF of the FIRST-undetected SDSS AGNs. When they compare this result to the CCF between 
FIRST-detected and undetected SDSS AGNs, they find a 2.5$\sigma$ difference in $r_0$,
using only the diagonal elements of the covariance matrix.
Fixing the slope and not using the full covariance matrix likely increases the systematic 
uncertainties in their results.  By contrast, our measurements test a much narrower 
redshift range, use the full covariance matrix, allow free $r_0$ and $\gamma$ values 
for the power law fits, and use the HOD modeling approach. \cite{shen_strauss_2009}
test much brighter AGN samples at higher redshifts than our samples here.  The fact 
that they find a difference in the clustering of radio-detected versus non-detected AGNs 
and we do not may be a consequence of the samples having different luminosities or redshifts.

If broad-line AGNs at $z<0.5$ indeed have no clustering dependence on radio emission,
this may suggest that different physical 
mechanisms trigger radio emission in broad-line and narrow-line AGNs, as narrow-line AGNs do 
have a clustering dependence with radio emission (\citealt{mandelbaum_li_2009}) at fixed 
stellar mass. This would be somewhat surprising as significant radio emission in 
all AGNs is believed to be related to jet phenomena (e.g., \citealt{blandford_payne_1982}).

\subsection{Luminosity Dependence of the Clustering Signal}
\label{lumdependence}
In papers I and II, we reported a possible X-ray luminosity-dependence in the 
AGN clustering strength of RASS/SDSS AGN in the redshift range $0.16<z<0.36$
(see also Fig.~\ref{xray_optical_b_HOD}, green and yellow green symbols).  
The exclusion of radio-detected RASS/SDSS AGNs performed here tests whether 
the weak  $L_{\rm X}$-dependence of AGN clustering observed is due to the presence 
of radio-loud AGNs in the higher $L_{\rm X}$ sample (see Sect.~\ref{radio_RASS}).

As shown in paper II, this luminosity-dependent clustering  is
more significant in the one halo term than in the two halo term. Using only
larger scales of $r_P>0.7$ $h^{-1}$ Mpc for the HOD modeling in this
paper (in order to derive bias parameter in a consistent way for all
AGN samples) increases the error on the individual measurements. This
results in a decrease in the significance of the luminosity-dependent
clustering. We find that the 1.5$\sigma$ (2.0$\sigma$ using
the power law bias of this paper) clustering difference between the
lower and higher $L_{\rm X}$ AGN samples remains roughly constant when we
exclude radio-detect AGNs from both subsamples (1.7$\sigma$ using the HOD bias;
2.4$\sigma$ using the power law bias of this paper).

The difference in the clustering strength of the lower and higher $L_{\rm X}$ 
RASS/SDSS AGN samples in the redshift range of $0.07<z<0.16$ is 1.2$\sigma$ (HOD model).
The error bars on the bias of these samples are high, 
due to the low number of objects. The significance of the individual redshift 
measurements do not provide strong evidence for an X-ray luminosity dependence of the 
AGN clustering strength. 

Other studies find similar weak trends at low redshifts (Fig.~\ref{lowz_b}). 
\cite{cappelluti_ajello_2010} compute the clustering signal for {\em Swift}/BAT 
15--55 keV selected AGNs and find a clustering difference of 1.6$\sigma$ for 
the higher $L_{\rm X}$ AGNs relative to the lower $L_{\rm X}$ 
AGNs. Comparing the photometric galaxy density around spectroscopic 
SDSS AGNs, \cite{serber_bahcall_2006} and \cite{strand_brunner_2008} 
find that higher luminosity AGNs have more overdense galaxy environments 
compared to lower luminosity AGNs at scales smaller than 0.5 Mpc. These studies 
are not directly measuring AGN clustering, rather they focus on the immediate 
environments of AGNs. 

We do not detect an optical luminosity-dependence of the RASS/SDSS AGN clustering.
Given the large sample size and low resulting errors in these samples, this is a
constraining result. 
Previous clustering measurements of optically-selected AGNs using the 
2dF QSO Redshift Survey (2QZ, \citealt{boyle_shanks_2000}), 2dF-SDSS LRG and QSO 
(2SLAQ, \citealt{cannon_drinkwater_2006}), and SDSS also find little evidence for 
an optical luminosity dependence of the AGN clustering strength (e.g., 
\citealt{croom_boyle_2002}; \citealt{angela_shanks_2008}; \citealt{mountrichas_sawangwit_2009}).
\cite{mountrichas_sawangwit_2009} use CCFs between AGNs and LRGs and find
some indication that bright SDSS AGNs cluster less than faint 2SLAQ QSOs, although
the result is only marginally significant (1.6$\sigma$).
On the other hand, \cite{shen_strauss_2009}
detect a stronger clustering strength for the 10\% most luminous SDSS DR5 AGNs 
($0.4\lesssim z \lesssim 2.5$) at the $\sim$2$\sigma$ level. However, they caution 
that the dynamical range in luminosity probed is narrow and the sample size in the 
luminosity subsamples not large enough to yield constraining clustering measurements. 
\cite{porciani_norberg_2006} suggest that a luminosity dependence of the clustering may 
be more evident at $z>1.3$, while \cite{angela_shanks_2008} find hints that the lower 
redshift ranges ($z<1.3$) may show more dependence with luminosity.

The optical and X-ray luminosities of broad-line AGNs are connected via the 
optical--to--X-ray spectral index, which measures the ratio of the rest-frame 
luminosity density at 2500 \AA\ to 2 keV. 
Although the relation has some scatter (see \citealt{anderson_margon_2007} 
for the optical--to--X-ray spectral index for RASS/SDSS AGNs), more X-ray luminous
AGNs are, on average, also intrinsically brighter in the optical. If indeed there is 
a weak X-ray luminosity dependence of the AGN clustering strength,  
the very narrow $M_i$ range that the optical SDSS AGNs span would hamper a detection. 
Identifying AGN activity using X-ray emission allows us to identify low-luminosity AGNs, 
where the optical light is dominated by the host galaxy. This results in a wider luminosity 
range in X-ray emission than in the optical (Fig.~\ref{O_SDSS_radio_X-ray}, lower panel) 
in the two lowest redshift ranges.
At $0.16 <z<0.36$, the mean optical luminosity of the high $M_i$ SDSS AGN sample is 
only a factor 2 higher (0.75 mag) than the low $M_i$ SDSS AGN sample, while 
the X-ray luminosities between the high and low $L_{\rm X}$ samples differ 
by a factor of 4.4. This is also seen when considering the luminosity difference between the 
90th percentile in the high luminosity samples and the 10th percentile in the low 
luminosity samples (factor $f =4.3$ for optical SDSS AGNs, $f= 17.8$ for RASS/SDSS 
AGNs) and covered luminosity range ($f=30$ for optical SDSS AGNs, $f=310$ for 
RASS/SDSS AGNs).

Hence, to detect a possible optical luminosity dependence in the broad-line 
AGN clustering strength, as might be expected if there is a weak X-ray luminosity 
dependence, considering the optical--to--X-ray luminosity relation, a wider optical luminosity 
range has to be tested. As these samples already include the brightest objects, this 
is only possible if one can include lower optical luminosities where broad-line AGNs are 
not effectively selected due to an increase in the host galaxy starlight fraction.
Only much deeper surveys at larger redshifts may yield the dynamical range to test 
the luminosity dependence for optically selected broad-line AGNs.

The possible X-ray luminosity dependence of broad-line AGN clustering detected at low redshifts 
(in that more X-ray luminous AGNs are more clustered than less X-ray luminous AGNs) may be difficult
to reconcile with the result that, on average, low luminosity (mainly narrow-line)
X-ray selected AGNs at higher redshifts are more clustered than luminous optically-selected 
broad-line AGNs (see Fig.~\ref{xray_optical_b}). If the X-ray luminosity dependence 
is real, this may suggest that different physical processes trigger AGN activity at different 
cosmological times or at different luminosities.

\subsection{Current Picture of AGN Clustering}
Although AGN clustering measurements are currently not as constraining and 
the interpretation of the results is not as clear as for galaxy clustering 
measurements, some general findings have emerged in the last few years.

At low redshifts ($z\la0.5$), broad and narrow-line AGN cluster similarly to 
inactive galaxies, occupying DMH masses of 
log $(M_{\rm DMH}/[h^{-1}\,M_{\odot}]) \sim 12.0-13.5$. 
This DMH mass range includes cases where the DMH is dominated by one 
$\gtrsim$$L^*$ galaxy or a small galaxy group composed of multiple such 
galaxies (\citealt{zehavi_zheng_2005,zehavi_zheng_2011}).

Independent of the selection method, the clustering strength of broad-line AGNs 
does not significantly change, while narrow-line AGNs show a significant increase 
in the clustering amplitude when radio-selected narrow-line AGNs are studied. 
Finally, more X-ray luminous broad-line AGNs may cluster more strongly than their 
lower luminosity counterparts. Although the various AGN samples have 
different luminosities, radio-loud, optically-selected 
narrow-line AGNs, very luminous X-ray AGNs, and red galaxies reside in somewhat 
similar high DMH masses. Lower luminosity X-ray AGNs, optical narrow-line 
AGNs with no radio emission, and blue galaxies tend to be found in lower DMH masses.

At high redshifts ($z\ga0.7$), X-ray selected AGN samples appear to cluster more strongly 
than optically-selected AGNs. The reason for this remains unclear. Possibly either 
low-luminosity or narrow-line (X-ray absorbed) AGNs cluster more strongly than very luminous 
broad-line optical AGNs. Additionally, as some of the X-ray clustering studies significantly 
underestimate their systematic uncertainties it may turn out that these measurements are 
consistent with optical AGN clustering measurements. More high-z AGN clustering measurements 
based on larger samples are needed to gain a clearer picture.

In this paper we use AGN samples based on SDSS. In the very near future there is not 
another planned survey that includes photometry and a dedicated extensive spectroscopic follow-up 
program for AGNs over such large areas as that covered by SDSS. 
As both large co-moving volumes and spectroscopic redshifts are essential for precise 
AGN clustering measurements, major improvement in our X-ray and optically-selected low redshift 
AGNs are therefore not expected in the very near future. BOSS (\citealt{eisenstein_weinberg_2001}) 
and BigBOSS (\citealt{schlegel_abdalla_20011}) will detect high redshift AGNs 
at $z \sim 2.2$, which will improve AGN clustering measurements at higher redshifts. 

In the coming years, the {\em ROSAT} successor {\em eROSITA} 
(\citealt{predehl_andritschke_2007}) will perform an all-sky survey in the hard 
and soft X-rays, probing much fainter than RASS, which is expected to detect up to $\sim$3 
million AGNs. Additionally, the Large Synoptic Survey Telescope (LSST, \citealt{ivezic_tyson_2008}) 
is expected to identify $\sim$2 million AGNs in optical bands.
{\em eROSITA} and LSST have the potential to significantly improve AGN clustering measurements 
at low and high redshifts, though only if there are dedicated large spectroscopic 
follow-up programs.


\section{Conclusions}
This work presents AGN clustering measurements at low redshifts in three independent redshift 
ranges: $0.07<z<0.16$, $0.16<z<0.36$, and $0.36<z<0.50$. Extending the use of the cross-correlation method of 
\cite{krumpe_miyaji_2010}, we infer the auto-correlation for both 
{\it ROSAT} All-Sky Survey (RASS) and optically-selected SDSS broad-line AGNs.
As tracer sets we use SDSS main galaxies, SDSS luminous 
red galaxies, and very luminous red galaxies (extended LRG sample). 
We apply the HOD model method (\citealt{miyaji_krumpe_2011}) 
directly to the measured CCFs to derive the bias parameter.  
We study the impact of different AGN selections on the clustering signal of 
broad-line AGNs, i.e., by excluding radio-detected AGNs. Furthermore, we compute 
the clustering strength for RASS-selected narrow-line AGNs. 

We find no statistically convincing difference in
the clustering of X-ray and optically-selected 
broad-line SDSS AGNs at low redshifts ($z<0.5$).
Different AGN selections based on either X-ray (RASS), 
optical (SDSS), or radio (FIRST), and combinations of these, 
do not significantly change the clustering 
signal for broad-line AGNs. 
This appears to be in contrast to other studies that find stronger clustering 
for radio-loud AGNs (e.g., \citealt{mandelbaum_li_2009}).
However, those results are based on narrow-line, low 
luminosity AGNs, while our sample consists of more luminous broad-line AGNs.
For the X-ray selected broad-line RASS/SDSS AGNs we find HOD bias values of 
1.23$^{+0.09}_{-0.08}$, 1.30$^{+0.09}_{-0.08}$, and 1.02$^{+0.14}_{-0.09}$ in the 
redshift ranges $0.07<z<0.16$, $0.16<z<0.36$, and $0.36<z<0.50$, respectively, 
while the HOD bias values for the optically selected broad-line SDSS AGNs 
are 0.95$^{+0.17}_{-0.10}$, 1.29$^{+0.05}_{-0.05}$, and 1.33$^{+0.07}_{-0.08}$, respectively.
The corresponding inferred typical dark matter 
halo masses hosting our broad-line AGNs are in the range of 
log $(M_{\rm DMH}/[h^{-1}\,M_{\odot}]) \sim 12.5-13.2$
and are consistent with those occupied 
by $\gtrsim$$L^*$ galaxies at these redshifts.

We measure the clustering of RASS selected narrow-line AGNs, which 
consists of a mix of NLS1s, Seyfert 1.5, 1.8, 1.9, and Seyfert 2 candidates.
We do not find a significantly lower clustering amplitude of RASS 
narrow-line AGNs compared to broad-line AGNs, although these measurements are 
subject to large uncertainties. In addition, we rule out that RASS narrow-line AGNs 
cluster significantly more strongly than broad-line AGNs at low redshifts. 

We show that the exclusion of radio-detected RASS/SDSS AGNs in $0.16<z<0.36$ does not 
change the weak X-ray luminosity dependence of the AGN clustering strength that 
we find in paper I (in that higher $L_{\rm X}$ AGNs cluster more strongly than lower $L_{\rm X}$ AGNs 
at $\sim2\sigma$). 
We do not detect an optical luminosity dependence of the broad-line AGN clustering 
in the same redshift range, though this result is not particularly constraining 
due to the narrow $M_i$ range that is covered.

We derive the bias parameter based on the best power law fit, the standard 
method used in literature, as well as by using HOD modeling. Important differences 
between the two techniques are found for some AGN samples.  In particular, 
using a power law fit can underestimate the bias compared to HOD modeling. 
We show that HOD model bias parameters 
are more reliable and more accurately reflect the large-scale clustering strength. 
Larger AGN samples will be provided by future missions such as 
{\em eROSITA}.  As these samples will have lower statistical uncertainties, HOD 
model bias parameters should be used to avoid introducting systematic errors 
that could exceed the statistical errors and thus possibly lead to a 
misinterpretation of clustering measurements.


\acknowledgments
We would like to thank Richard Rothschild, Alex Markowitz, Slawomir Suchy, and
Stephen Smith for helpful discussions. Furthermore, we thank Ryan Hickox for providing the 
inferred ACFs of their samples. Last but not least, we thank the referee for a very 
helpful report.

This work has been supported by NASA grant NNX07AT02G,  CONACyT Grant Cient\'ifica B\'asica 
\#83564, UNAM-DGAPA Grants PAPIIT IN110209 and IN109710.

The {\em ROSAT} Project was supported by the Bundesministerium f{\"u}r Bildung 
und Forschung (BMBF/DLR) and the Max-Planck-Gesellschaft (MPG).
Funding for the Sloan Digital Sky Survey (SDSS) has been 
provided by the Alfred P. Sloan Foundation, the Participating 
Institutions, the National Aeronautics and Space Administration, 
the National Science Foundation, the U.S. Department of Energy, 
the Japanese Monbukagakusho, and the Max Planck Society. 
The SDSS Web site is http://www.sdss.org/.

The SDSS is managed by the Astrophysical Research Consortium (ARC) 
for the Participating Institutions. The Participating Institutions 
are The University of Chicago, Fermilab, the Institute for Advanced 
Study, the Japan Participation Group, The Johns Hopkins University, 
Los Alamos National Laboratory, the Max-Planck-Institute for 
Astronomy (MPIA), the Max-Planck-Institute for Astrophysics (MPA), 
New Mexico State University, University of Pittsburgh, Princeton 
University, the United States Naval Observatory, and the 
University of Washington. 

This research also made use of computing facility 
available from Departmento de Superc\'omputo, DGSCA, UNAM.



\end{document}